\documentclass[10pt,conference,letterpaper]{IEEEtran}
\usepackage{graphicx}
\usepackage{amssymb}
\usepackage{url}
\usepackage[dvipsnames]{xcolor}
\usepackage{mypackage}
\providecommand{\Description}[1]{}

\usepackage{booktabs}
\usepackage{multirow}
\usepackage{colortbl}
\usepackage{tabularray}
\UseTblrLibrary{booktabs}

\usepackage{tikz}
\usetikzlibrary{calendar,fpu}
\tikzset{
    moon colour/.style={
        moon fill/.style={
            fill=#1
        }
    },
    sky colour/.style={
        sky draw/.style={
            draw=#1
        },
        sky fill/.style={
            fill=#1
        }
    },
    southern hemisphere/.style={
        rotate=180
    }
}
\usepackage[hidelinks,bookmarksnumbered=true,bookmarksopen=true]{hyperref}

\makeatletter
\pgfcalendardatetojulian{2010-01-15}{\c@pgf@counta} 
\def\synodicmonth{29.530588853}
\newcommand{\moon}[2][]{%
    \edef\checkfordate{\noexpand\in@{-}{#2}}%
    \checkfordate%
    \ifin@%
        \pgfcalendardatetojulian{#2}{\c@pgf@countb}%
        \pgfkeys{/pgf/fpu=true,/pgf/fpu/output format=fixed}%
        \pgfmathsetmacro\dayssincenewmoon{\the\c@pgf@countb-\the\c@pgf@counta-(7/24+11/(24*60))}%
        \pgfmathsetmacro\lunarage{mod(\dayssincenewmoon,\synodicmonth)}
        \pgfkeys{/pgf/fpu=false}
    \else%
        \def\lunarage{#2}%
    \fi%
    \pgfmathsetmacro\leftside{ifthenelse(\lunarage<=\synodicmonth/2,cos(360*(\lunarage/\synodicmonth)),1)}%
    \pgfmathsetmacro\rightside{ifthenelse(\lunarage<=\synodicmonth/2,-1,-cos(360*(\lunarage/\synodicmonth))}%
    \tikz [moon colour=white,sky colour=black,#1]{
        \draw [moon fill, sky draw] (0,0) circle [radius=1ex];
        \draw [sky draw, sky fill] (0,1ex)
            arc (90:-90:\rightside ex and 1ex)
            arc (-90:90:\leftside ex and 1ex)
            -- cycle;
    }%
}

\AtBeginDocument{%
  }

\begin{document}

\title{\textbf{\textcolor{TitleBlue}{\textsf{LUCID}}}: An Agentic AI Framework on Digital-Twin in the Loop for QoS-Guaranteeing Robotic Control}
\author{
Hyeonsu Lyu$^{1}$,
Minwoo Kim$^{2}$,
Sehyun Ryu$^{2}$,
Hyun Jong Yang$^{1}$%
\thanks{$^{1}$Hyeonsu Lyu and Hyun Jong Yang are with the
Department of Electrical and Computer Engineering and the
Institute of New Media and Communications,
Seoul National University, Seoul, Republic of Korea.}%
\thanks{$^{2}$Minwoo Kim and Sehyun Ryu are with
Pohang University of Science and Technology (POSTECH),
Pohang, Republic of Korea.}%
}
\IEEEoverridecommandlockouts
\IEEEaftertitletext{%
    \begin{center}
        \includegraphics[width=0.8\textwidth]{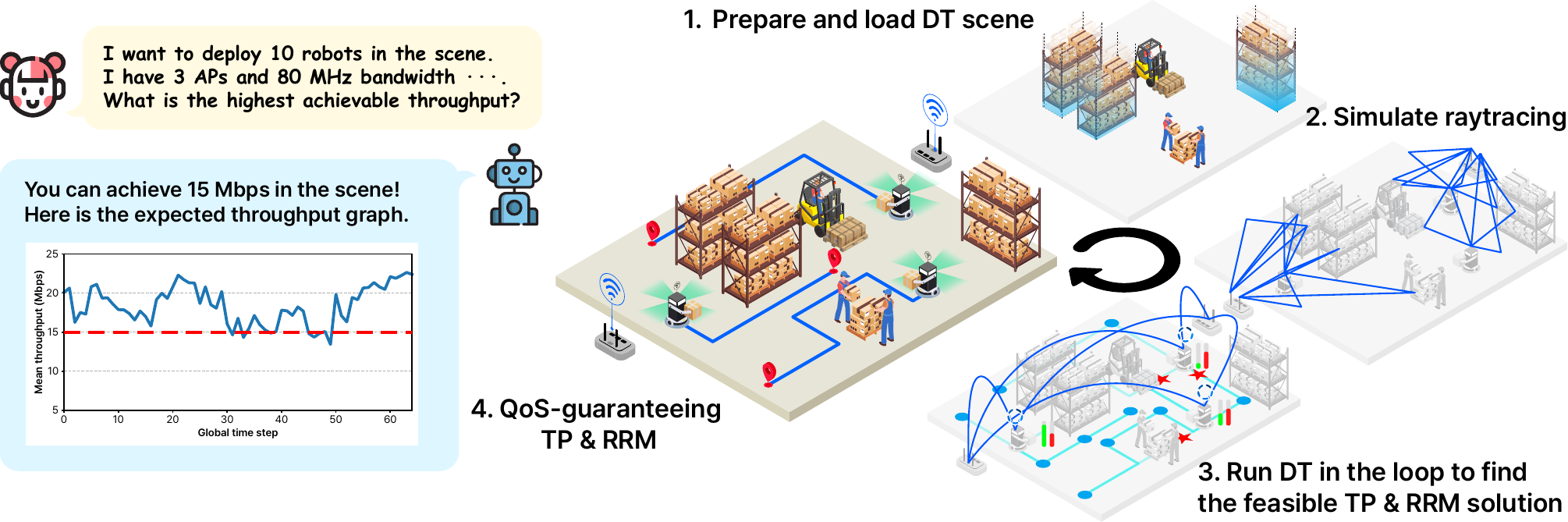}\\
        \refstepcounter{figure}\label{fig:lucid system model}%
        \footnotesize\textbf{Fig.~\thefigure.} Conceptual illustration of \lucid and automated agent.
    \end{center}
}
\maketitle

\begin{abstract}
Cloud robotics relies on the timely uplink of high-volume sensing streams, yet dynamic environments continually shift the feasible combinations of trajectories, active-robot count, and per-robot QoS.
Because existing approaches formulate trajectory planning (TP) and radio resource management (RRM) as a single fixed optimization problem, they cannot reconfigure these coupled decisions as conditions evolve, resulting in transient QoS violations.
However, evolving operator intents change which quantities-such as the active-robot count and per-robot QoS-are fixed, optimized, or relaxed.
Furthermore, the computational cost of evaluating trajectory-dependent wireless conflicts has made it difficult to build large-scale Digital-Twin-in-the-Loop (DITL) testbeds responsive enough for such dynamic orchestration.
We present \lucid, an LLM-agent--orchestrated, uplink-aware cloud-robotics pipeline that moves TP--RRM from solving a fixed formulation to dynamically orchestrating optimization problem schemas within a DITL environment.
Driven by the operator's high-level intent, \lucid treats the TP--RRM formulation as a bounded template whose variables, objectives, and constraints are dynamically configured, while SimBridge enables repeated ray-tracing evaluation by converting large-scale robotics scenes into wireless-ready DTs. 
By integrating collision-free path planning with a spectral-radius RRM validator, \lucid identifies wireless bottlenecks and restructures the problem schema on the fly to efficiently find the verified feasible state.
Experiments confirm that \lucid robustly adapts to changing intents, active-robot counts, and scenes, while a multimodal surrogate model, FastConfigNet, reduces planning latency.
\end{abstract}




\section{Introduction}
\label{sec:intro}

Cloud robotics increasingly relies on edge servers to process high-rate sensory streams from groups of autonomous mobile robots (AMRs) \cite{Kehoe15-TASE,Mohanarajah15-TASE}.
In warehouse environments, this architecture enables centralized perception and coordinated trajectory planning (TP), but timely uplink delivery over a shared and spatially varying wireless network remains vulnerable to transient blockages and multipath fading \cite{Talli23-INFOCOM,Lyu23-TII}.
Before deploying AMRs, an operator must therefore determine which combinations of the number of AMRs, per-AMR uplink quality-of-service (QoS), and TP are supportable under the available radio resources.
The resulting challenge is to identify feasible operating points and, when a desired point is infeasible, determine whether the AMR count, the QoS requirements, or TP should be adjusted.

\begin{figure}[b]
    \centering
    \includegraphics[width=\linewidth]{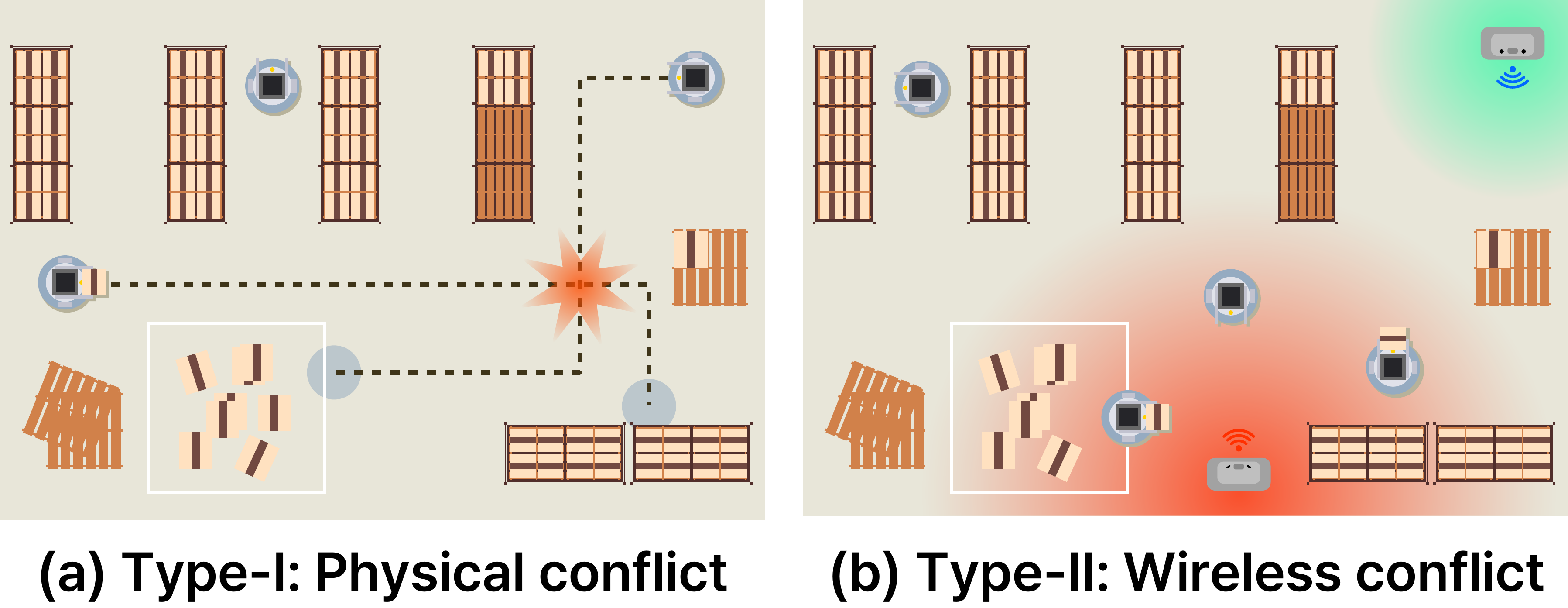}
    \caption{Two possible conflict types in warehouse environments.}
    \Description{Two side-by-side diagrams illustrating Type-I physical collisions and Type-II wireless conflicts in warehouse AMR operations.}
    \label{fig:collision_types}
\end{figure}

A fundamental difficulty is that TP and radio resource management (RRM) cannot be performed independently \cite{Lyu25-TWC_DBSPF}.
As illustrated in Fig.~\ref{fig:collision_types}(a), trajectories planned independently for individual AMRs may intersect in space and time, resulting in Type-I physical conflicts.
Conventional centralized multi-agent path planning can resolve these geometric conflicts by jointly coordinating the AMR trajectories \cite{Sharon12-AAAI,Barer21-SoCS}.
However, because conventional planners do not account for the radio resources required along those trajectories, the resulting collision-free plan may concentrate multiple AMRs in a radio-limited region.
Their simultaneous uplink QoS requirements may then become infeasible, resulting in the Type-II wireless conflict shown in Fig.~\ref{fig:collision_types}(b).
Thus, TP cannot be finalized before wireless feasibility is evaluated, and RRM cannot be applied merely as a subsequent step to fixed trajectories.

\begin{figure}[htb]
    \centering
    \includegraphics[width=\linewidth]{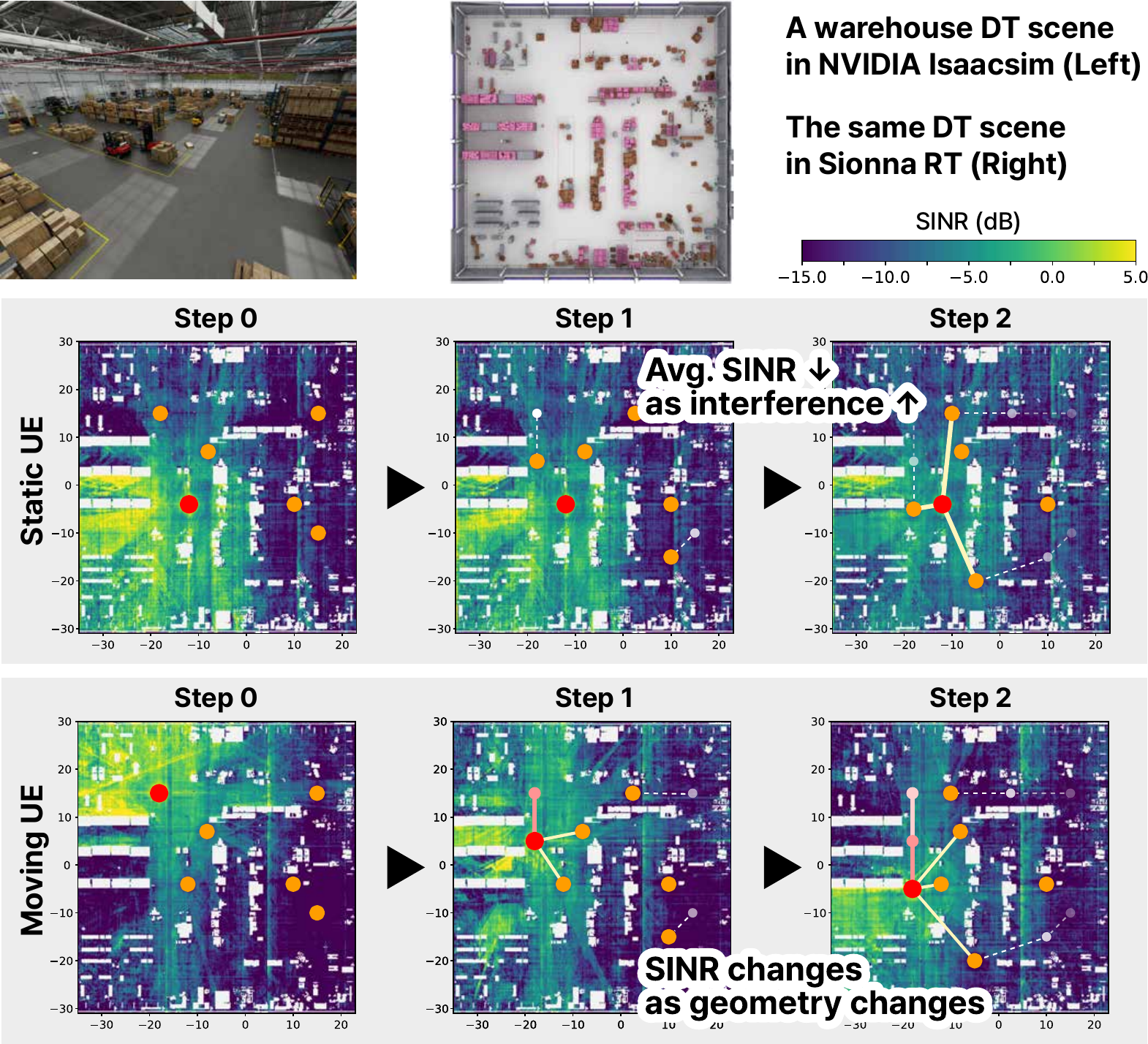}
    \caption{Uplink SINR map of two UEs over three steps. The heatmap shows the SINR of the red UE when the orange UEs act as interferers.}
    \Description{A visualization of SINR evolution for warehouse robots in an NVIDIA Isaac Sim scene and its Sionna RT digital twin.}
    \label{fig:dt_demo}
\end{figure}

Evaluating TP--RRM feasibility is itself challenging because wireless conditions evolve over both space and time \cite{Boban14-TVT}.
Fig.~\ref{fig:dt_demo} illustrates how wireless feasibility changes as multiple AMRs move and communicate concurrently in the NVIDIA PhysicalAI SimReady Warehouse-01 environment \cite{NVIDIA-SimreadyWarehouse}.
Their relative positions and simultaneous transmissions continually alter the interference pattern, while their motion changes blockage and multipath propagation along the trajectories \cite{Zubow24-Arxiv}.
These interactions can produce short-lived but severe channel degradations at specific locations and time slots; such transient events are particularly important in deadline-driven communication and control systems \cite{Popovski19-TCOM,Ho24-TNSM}.
Conventional stochastic channel models based primarily on large-scale statistical properties may smooth out such localized events and thereby miss transient QoS violations \cite{Ropitault25-Arxiv}.
Scene-specific electromagnetic (EM) simulation and ray tracing can instead provide location-dependent channel estimates along candidate trajectories \cite{Alkhateeb19-deepmimo,Guneser25-VTC}.
Accurately identifying the feasible operating region therefore requires a scene-specific wireless digital twin \cite{Khan22-COMST} that resolves the channel and interference conditions along the joint AMR trajectories and evaluates them with the corresponding RRM decisions \cite{Pegurri25-INFOCOM,SionnaSYSmeetsRT}.

However, existing large-scale robotics scenes are not directly usable as wireless digital twins.
The robotics scene considered in this work, for example, is designed primarily for visual rendering and physical simulation \cite{NVIDIA-SimreadyWarehouse}.
Such scenes commonly lack EM material descriptions while retaining geometric details that make repeated ray-tracing evaluation unnecessarily expensive.
In contrast, contemporary wireless ray-tracing toolchains require meaningful material assignments \cite{Zhang24-WisegRT} and computationally tractable scene representations.
Directly processing unmodified robotics scenes for repeated ray-tracing evaluation can therefore introduce substantial computational overhead, preventing their use within a responsive feasibility-search loop \cite{Zubow24-Arxiv}.
A practical solution must transform large-scale robotics scenes into wireless-ready representations while preserving the propagation features needed to evaluate trajectory-dependent QoS violations.

Even with such an evaluation infrastructure, characterizing the feasible operating region remains computationally difficult.
Each candidate operating point requires trajectory-dependent DT evaluation \cite{Ropitault25-Arxiv} and coupled TP--RRM optimization \cite{Lyu25-TWC_DBSPF}, making the search increasingly expensive as the number of AMRs grows.
Moreover, the operator's request does not merely change a numerical input to a fixed solver.
Different priorities change which variables are fixed, which are optimized, and which constraints may be relaxed.
Maximizing per-AMR QoS for a fixed number of AMRs, for example, requires a different optimization formulation from maximizing the supported number of AMRs under a fixed QoS requirement.
When a requested point is infeasible, the formulation must further reflect whether the operator prefers rerouting AMRs, reducing sensing fidelity, or decreasing the number of active AMRs.
Recent work has shown that LLMs can interface with structured networking inputs and tools \cite{Wu24-SIGCOMM-NetLLM}, while also highlighting the need for explicit safeguards that ensure validity and reliability.
Feasibility search must therefore be conditioned on operator intent, with each search step translated into a structured and verifiable TP--RRM problem \cite{Wang24-ProcACMNetw}.

To address these challenges, we present \lucid, an LLM-assisted, uplink-aware cloud-robotics pipeline operating in a digital-twin-in-the-loop (DITL) environment for warehouse operations.
\lucid incorporates an automated scene-processing module, which we call SimBridge, to transform large-scale robotics scenes into wireless-ready DTs suitable for repeated trajectory-dependent evaluation.
The LLM component translates operator priorities into structured TP--RRM optimization problems within predefined formulation schemas and coordinates the relevant DT, TP, and RRM modules throughout the feasibility search.
Moreover, \lucid stores previously evaluated intents, configurations, and feasibility outcomes in a data lake and uses them to train FastConfigNet, a multimodal surrogate model that rapidly screens candidate configurations before costly DT-based TP--RRM evaluation.
Together, these components enable \lucid to identify feasible operating points defined by the number of AMRs and per-AMR QoS and to adapt the search efficiently to changing operator requirements.

\section{Intent-Conditioned TP--RRM System Model}

\begin{figure}[htb]
    \centering
    \includegraphics[width=\linewidth]{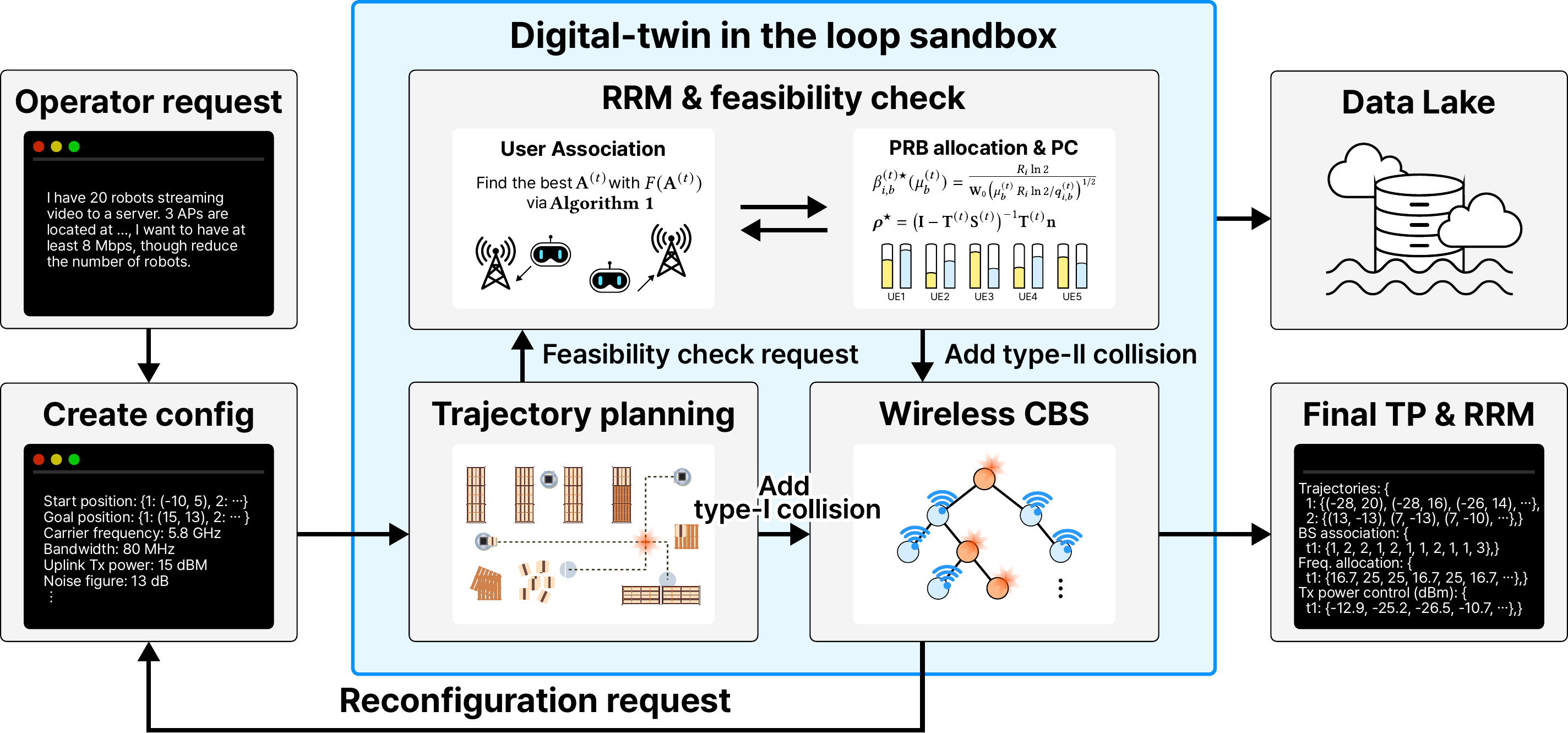}
    \caption{DITL workflow for intent-conditioned TP--RRM evaluation.}
    \Description{An operator request is converted into a candidate configuration. The configuration and DT environment instantiate trajectory-planning and RRM feasibility modules. A solver-verified evaluation returns trajectories and radio-resource variables; a validator failure leads to a reconfiguration request. Each evaluation and its outcome are stored in the data lake.}
    \label{fig:02_DITL_overview}
\end{figure}

Figure~\ref{fig:02_DITL_overview} shows the DITL workflow used to evaluate an operator request.
The request is translated into a candidate operating-point configuration, which instantiates a TP--RRM problem template in the digital-twin (DT) environment.

In the \emph{Create config} stage, the agent maps the request to a candidate configuration.
Together with the DT environment, the configuration specifies one instance of the TP--RRM template.
The trajectory planner and RRM validator then either return feasible trajectories and allocations or identify the failure that motivates the next configuration.
The agent selects and revises configurations, while the planning and validation modules perform the corresponding feasibility evaluation.

\myparagraph{Intent-Conditioned TP--RRM Problem Template}
We denote the DT environment used for evaluation as $\mathsf{env}$.
It combines the materialized and geometry-reduced scene produced by SimBridge with the deployment-specific BS layout and the ray-tracing results needed to query $h_{i,b}^{(t)}$ along a candidate trajectory.
For an operator request $\mathbf u$, $\mathsf{cfg}(\mathbf u)$ denotes the system parameters of the candidate operating point evaluated in this environment.

The centralized edge server plans robot motion and uplink scheduling over discrete time slots.
We index the $I$ robots, BSs, obstacles, and time slots by $\mathcal{I}$, $\mathcal{B}$, $\mathcal{K}$, and $\mathcal{T}$, respectively, with slot duration $\Delta t$.
Robot $i$ has position $\mathbf{x}_i^{(t)}\in\mathbb R^2$, initial position $\mathbf{x}_i^{(0)}$, maximum speed $v_{\max}$, and destination $\mathbf{x}_i^{\mathrm{goal}}$.
The position of obstacle $k$ at slot $t$ is $\mathbf{o}_k^{(t)}$.
The configuration $\mathsf{cfg}(\mathbf u)$ collects the active sets $\mathcal I$, $\mathcal B$, $\mathcal K$, and $\mathcal T$; the slot duration; robot start and goal positions; $v_{\max}$ and $d_{\min}$; the rate requirements $\{R_i\}$; the resource limits $B$ and $\rho_{\max}$; and the noise density $N_0$.
Robot motion must satisfy
\begin{align}
    &\|\mathbf{x}_i^{(t+1)}-\mathbf{x}_i^{(t)}\|
    \le v_{\max}\Delta t,~
    \mathbf{x}_i^{(T)}=\mathbf{x}_i^{\mathrm{goal}},~\forall i,
    \label{eq:constraint_velocity}\\
    &\|\mathbf{x}_i^{(t)}-\mathbf{x}_j^{(t)}\|,
    \|\mathbf{x}_i^{(t)}-\mathbf{o}_k^{(t)}\|
    \ge d_{\min},~\forall i\neq j,k.
    \label{eq:constraint_safe_distance}
\end{align}
For user $i$, let $a_{i,b}^{(t)}\in\{0,1\}$ denote association with BS $b$, $\beta_{i,b}^{(t)}$ the allocated bandwidth, $\rho_i^{(t)}$ the transmit power spectral density (PSD), and $h_{i,b}^{(t)}$ the uplink channel.
With noise density $N_0$, throughput $\gamma_{i,b}^{(t)}$ is given by
\begin{align}
    \gamma_{i,b}^{(t)}&=\beta_{i,b}^{(t)}\log_2
    \biggl(1+
        \frac{\rho_i^{(t)}\|h_{i,b}^{(t)}\|^2}
        {{\displaystyle\sum\limits_{b'\neq b}\sum\limits_{i'\neq i}}
        a_{i',b'}^{(t)}\rho_{i'}^{(t)}
        \|h_{i',b}^{(t)}\|^2+N_0}
    \biggr).
    \label{eq:instant_rate}
\end{align}
Every active robot associates with one BS and must meet its required uplink rate $R_i$:
\begin{align}
    \sum_{b\in\mathcal B}a_{i,b}^{(t)}&=1,~\forall i,t;~
    \sum_{b\in\mathcal B}a_{i,b}^{(t)}\gamma_{i,b}^{(t)}\ge R_i,~\forall i,t.
    \label{eq:constraint_QoS}
\end{align}
The per-BS bandwidth and per-AMR PSD constraints are
\begin{align}
    \sum_{i\in\mathcal I}\beta_{i,b}^{(t)}&\le B,~\forall b,t;~
    0\le\rho_i^{(t)}\le\rho_{\max},~\forall i,t.
    \label{eq:constraint_power}
\end{align}

We use the weighted sum of total travel distance and uplink transmit power as the objective. With a fixed slot duration and a constant motion cost per unit distance,
this can be interpreted as energy minimization. Therefore, minimizing this objective favors short paths and low-power allocations and can extend battery-limited operating duration. The weight $\lambda$ sets the tradeoff between the two energy terms.
The QoS-constrained problem template is
\begin{subequations}\label{eq:configurable_problem}
\begin{flalign}
  \problem[&\mathsf{env},\mathsf{cfg}(\mathbf u)]:\!
    \min_{\mathbf{X}, \mathbf{A}, \mathbf{B}, \mathbf{P}}\! \sum_{i,t}\|\mathbf{x}_{i}^{(t+1)}-\mathbf{x}_i^{(t)}\|\!+\!\lambda\sum_{i,t}\rho_i^{(t)}
    \hspace{-5mm}&&
    \\
    &\text{s.t.}~\eqref{eq:constraint_velocity}-\eqref{eq:constraint_safe_distance}~\textbf{(TP Constr.)},~
    \eqref{eq:constraint_QoS}-\eqref{eq:constraint_power}~\textbf{(RRM constr.)} \hspace{-5mm}&&
    \label{p1:constraints}
\end{flalign}
\end{subequations}
for 
$\mathbf X=\{\mathbf{x}_i^{(t)}\}$,
$\mathbf A=\{a_{i,b}^{(t)}\}$,
$\mathbf B=\{\beta_{i,b}^{(t)}\}$, and 
$\mathbf P=\{\rho_i^{(t)}\}$.

Problem~\eqref{eq:configurable_problem} contains binary trajectory and association decisions as well as nonconvex rate constraints.
The solver generates candidate trajectories $\mathbf X$ and performs slot-wise RRM feasibility evaluation over $(\mathbf A,\mathbf B,\mathbf P)$.
Section~III describes the solver and its conflict feedback between these stages.

\myparagraph{Operator Intent to Operating-Point Configuration}
The operator intent determines which fields of $\mathsf{cfg}(\mathbf u)$ are fixed and which one is varied across candidate operating points.
For example, suppose an operator asks whether a fixed DT environment can provide a target per-robot rate to a requested number of AMRs.
The agent begins by evaluating the requested AMR count; if that operating point is infeasible, it reduces the number of AMRs according to its relaxation priority and reevaluates the resulting configuration.
In contrast, when the operator fixes the number of AMRs and instead seeks higher per-robot performance, the agent holds the AMR count fixed and searches for the largest supportable per-robot rate.
Throughout this process, the search direction, bounds, and relaxation priority are maintained as orchestration state, separately from each feasibility evaluation.

Each search point supplies a configuration admitted by the template.
Within the DITL workflow of Fig.~\ref{fig:02_DITL_overview}, wireless CBS submits candidate trajectories to the RRM \& feasibility check block, which performs slot-wise RRM validation using the corresponding DT-derived channel gains.
The RRM validator returns $(\mathbf X,\mathbf A,\mathbf B,\mathbf P)$ for a feasible evaluation; otherwise, it returns a failure mode and the responsible robots or time slots for selecting the next candidate.
Section~III details the deterministic validation and wireless-CBS feedback loop that realizes this workflow, Section~IV archives and screens evaluated instances, and Section~V describes how SimBridge constructs the scene representation used by $\mathsf{env}$.

\section{Wireless-Aware TP--RRM Solver}

Within the DITL sandbox shown in Fig.~\ref{fig:02_DITL_overview}, the RRM validator is the RRM \& feasibility check block for each $\problem[\mathsf{env},\mathsf{cfg}(\mathbf u)]$ instance.
As the feasibility-testing module for wireless conflict-based search (CBS), it receives a candidate trajectory and the corresponding DT-derived channel gains, and tests whether its allocation procedure can find a feasible radio-resource allocation.
If no such allocation is found, it returns conflict diagnostics to the trajectory planner.
We first define its input, allocation procedure, and diagnosis, and then describe how the planner uses this feedback for path planning.

\myparagraph{RRM Feasibility Validation}
Given candidate trajectories $\mathbf X$, the robot positions and predicted channel gains are fixed at each time slot.
The per-slot RRM validator solves
\begin{align}
    \min_{\mathbf{A}^{(t)},\mathbf{B}^{(t)},\mathbf{P}^{(t)}}~&
    \sum_{i\in\mathcal{I}}\rho_i^{(t)}
    ~\text{s.t.}~\textbf{(RRM constr.)}.
    \label{eq:RRM_subproblem}
\end{align}
The validator couples power control and frequency allocation with local user-association updates.

\myparagraph{Power control}
For fixed $\mathbf{A}^{(t)}$ and $\mathbf{B}^{(t)}$, user $i$ has a unique serving BS $b_i$, where $a_{i,b_i}^{(t)}=1$.
We define the interference gain from user $j$ to user $i$ as
$
\sigma_{i,j}^{(t)}\coloneq
\big(1-a_{j,b_i}^{(t)}\big)\|h_{j,b_i}^{(t)}\|^2
$,
which vanishes when $j$ shares $i$'s serving BS.
With $\tau_i^{(t)}\coloneq 2^{R_i/\beta_{i,b_i}^{(t)}}-1$, the rate constraints can be written as
\begin{align}
    \boldsymbol{\rho}^{(t)}
    &\succcurlyeq \mathbf{T}^{(t)}
    \big(\mathbf{S}^{(t)}\boldsymbol{\rho}^{(t)}+\mathbf n\big),
    \label{eq:matrixineq}
\end{align}
where $\mathbf n$ is the thermal-noise vector and
\begin{flalign}
    &\mathbf{T}^{(t)}\!\coloneq
    \mathrm{diag}\bigg(
    \frac{\tau_1^{(t)}}{\|h_{1,b_1}^{(t)}\|^2},\ldots,
    \frac{\tau_I^{(t)}}{\|h_{I,b_I}^{(t)}\|^2}\bigg),
    ~\mathbf{S}^{(t)}\!\coloneq[\sigma_{i,j}^{(t)}]_{i,j}.
    \hspace{-3mm}
    &&
\end{flalign}

If $\rho(\mathbf{T}^{(t)}\mathbf{S}^{(t)})<1$, the inverse of
$\mathbf I-\mathbf{T}^{(t)}\mathbf{S}^{(t)}$ is nonnegative because
$(\mathbf I-\mathbf{T}^{(t)}\mathbf{S}^{(t)})^{-1}
=\sum_{k=0}^{\infty}(\mathbf{T}^{(t)}\mathbf{S}^{(t)})^k\succeq\mathbf0$.
The componentwise least feasible power vector is therefore
\begin{align}
    \boldsymbol{\rho}^{\star(t)}
    &=\big(\mathbf I-\mathbf{T}^{(t)}\mathbf{S}^{(t)}\big)^{-1}
    \mathbf{T}^{(t)}\mathbf n,
    \label{eq:optimal_power}
\end{align}
which uniquely minimizes the sum-power objective for the fixed association and bandwidth allocation.
The allocation is feasible only if $\rho_i^{\star(t)}\le\rho_{\max}$ for every robot.

\begin{figure}[t]
    \centering
    \includegraphics[width=\linewidth]{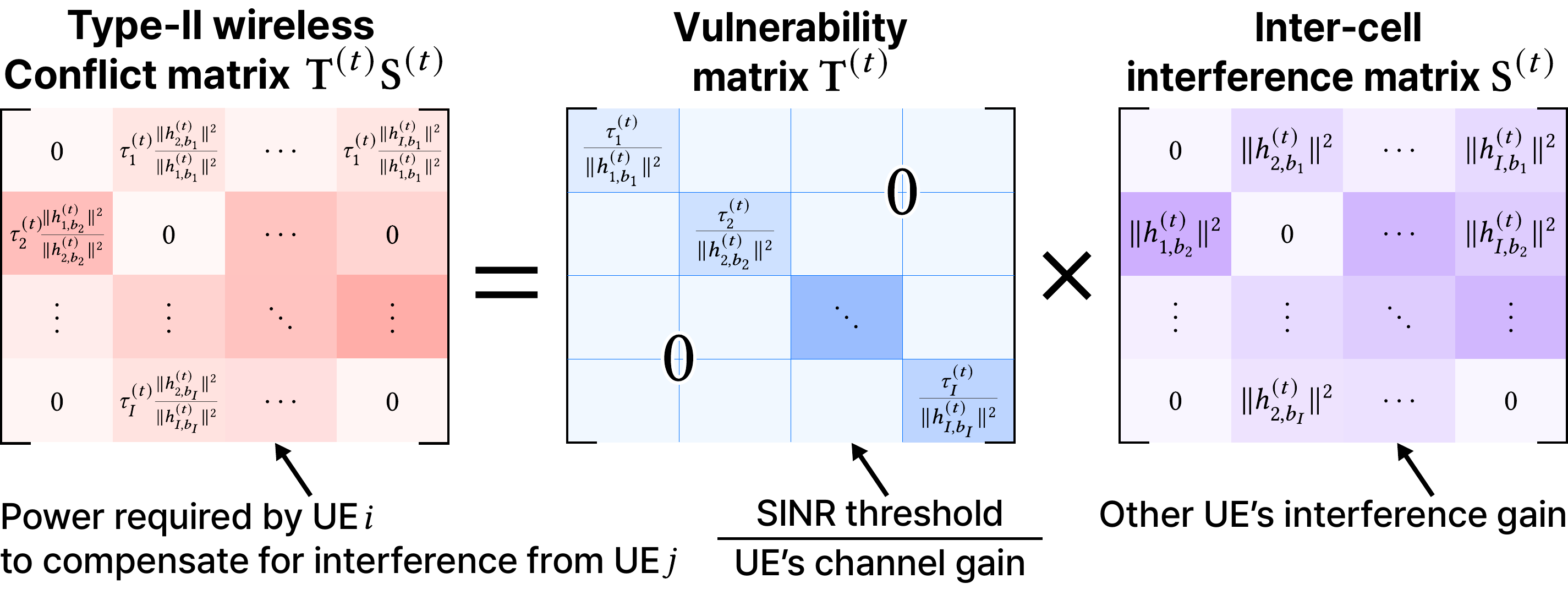}
    \caption{RRM coupling matrix for feasibility and conflict diagnosis.}
    \Description{A mathematical diagram illustrating the construction of the Type-II wireless conflict matrix through matrix multiplication. On the left side of the equals sign is the Type-II wireless conflict matrix, shaded in red with zeros on its main diagonal. On the right side, a diagonal vulnerability matrix is multiplied by an inter-cell interference matrix.}
    \label{fig:02_power_matrix_meaning}
\end{figure}

\myparagraph{Frequency allocation and user association}
For the current interference powers, define
\begin{align}
    q_{i,b}^{(t)}\coloneq \Big(
    {
    \sum_{b'\neq b}\sum_{i'\neq i}
    a_{i',b'}^{(t)}\rho_{i'}^{(t)}\|h_{i',b}^{(t)}\|^2+N_0}
    \Big)
    /{\|h_{i,b}^{(t)}\|^2}.
\end{align}
The rate constraint then lower-bounds the required power as
\begin{align}
    \rho_i^{(t)}\ge
    \sum_{b\in\mathcal B}a_{i,b}^{(t)}q_{i,b}^{(t)}
    \big(2^{R_i/\beta_{i,b}^{(t)}}-1\big).
    \label{eq:per_user_power_lower_bound}
\end{align}
Plugging in \eqref{eq:per_user_power_lower_bound} into problem~\eqref{eq:RRM_subproblem} gives the subproblem:
\begin{align}
    \!\!\min_{\mathbf{A}^{(t)},\mathbf{B}^{(t)}}~&
    \sum_{i,b}a_{i,b}^{(t)}q_{i,b}^{(t)}
    \big(2^{R_i/\beta_{i,b}^{(t)}}-1\big)
    ~\text{s.t.}~\textbf{(RRM constr.)}.
    \label{eq:ua_ra_subproblem}
\end{align}

For fixed $\mathbf A^{(t)}$ and $\{q_{i,b}^{(t)}\}$, the subproblem \eqref{eq:ua_ra_subproblem} is convex.
The KKT stationarity condition for user association then yields
\begin{flalign}
    &\hspace{19mm}
    2^{r_i/\beta_{i,b}^{(t)}}
    q_{i,b}^{(t)}r_i\ln2/(\beta_{i,b}^{(t)})^2
    =\mu_b^{(t)},
    \label{eq:stationarity_time}
    \\
    &\beta_{i,b}^{(t)\star}(\mu_b^{(t)})
    =\begin{cases}
    \displaystyle
    \frac{R_i\ln2}
    {2W_0\big(
    \sqrt{{\mu_b^{(t)}R_i\ln2}/{q_{i,b}^{(t)}}}/2\big)},
    &a_{i,b}^{(t)}=1,\\[-1.5mm]
    0,&a_{i,b}^{(t)}=0,
    \end{cases}
    \hspace{-5mm}&&
    \label{eq:optimal_beta}
\end{flalign}
where $W_0(\cdot)$ is the principal branch of the Lambert $W$ function.
The total allocated bandwidth is monotone in $\mu_b^{(t)}$, so a one-dimensional bisection selects the multiplier satisfying
$\sum_i a_{i,b}^{(t)}\beta_{i,b}^{(t)\star}=B$.

Because \eqref{eq:optimal_beta} determines the bandwidth allocation for a fixed association and the current interference powers, the remaining discrete step is to choose which BS serves each user.
This choice balances serving-link quality, per-BS bandwidth sharing, and inter-cell interference: associating a user with a stronger BS improves its desired link, but also changes the bandwidth available to other users at that BS and the interference observed across cells.
We initialize the search with the strongest-link association $b_i=\arg\max_b\|h_{i,b}^{(t)}\|^2$ and $K_{\mathrm{init}}=5$ random-restart associations, and select the feasible seed with the smallest evaluated objective.
The search then considers moving one user from its current BS to another BS.
Because only the memberships of these two BSs change, their bandwidth allocations are recomputed using \eqref{eq:optimal_beta} before updating the power vector and testing feasibility.
A reassignment is accepted when it remains feasible and decreases
\begin{align}
    F(\mathbf A^{(t)})\coloneq
    \sum_{i,b}a_{i,b}^{(t)}q_{i,b}^{(t)}
    \big(2^{R_i/\beta_{i,b}^{(t)\star}}-1\big).
    \notag
\end{align}
Over at most $L_{\max}$ updates, these moves refine the association toward a lower-power feasible configuration.

\myparagraph{Feasibility diagnostics}
Figure~\ref{fig:02_power_matrix_meaning} interprets
$\mathbf{T}^{(t)}\mathbf{S}^{(t)}$ as the interference-coupling matrix: increasing one user's power raises the compensating power required by other users.
The validator distinguishes two failure modes.
If $\rho(\mathbf{T}^{(t)}\mathbf{S}^{(t)})\ge1$, no finite power vector satisfies the SINR constraints for the current association and bandwidth allocation; we call this an \emph{interference burst}.
If the spectral-radius condition holds but some $\rho_i^{\star(t)}>\rho_{\max}$, the allocation has a \emph{power-cap violation}.

\begin{figure}[t]
    \centering
    \includegraphics[width=\linewidth]{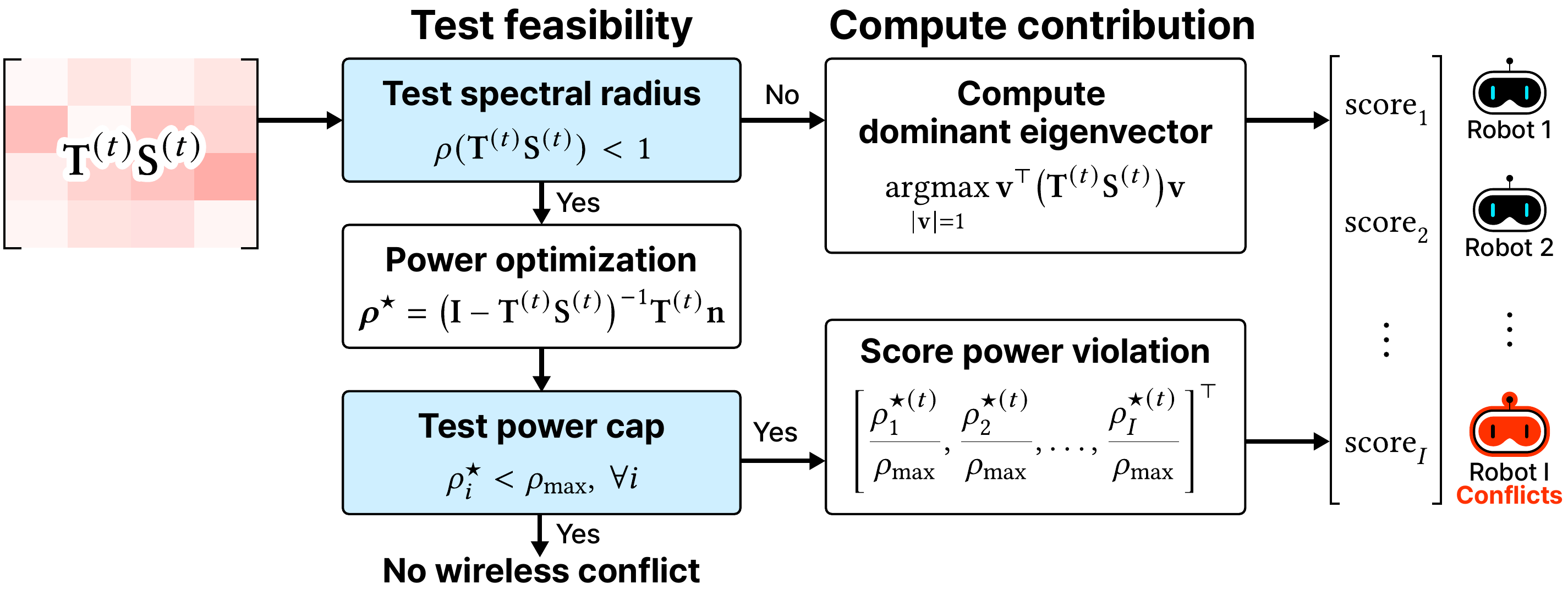}
    \caption{Per-slot RRM feasibility evaluation and conflict localization.}
    \Description{A flowchart that first tests the spectral radius and power cap. A failed spectral-radius test invokes dominant-eigenvector scoring, while a failed power-cap test scores the normalized power deficits. Both paths identify a culprit robot for trajectory replanning.}
    \label{fig:02_infeasibility_test}
\end{figure}

Figure~\ref{fig:02_infeasibility_test} shows how either failure is localized to a robot.
For an interference burst, the nonnegative matrix $\mathbf{T}^{(t)}\mathbf{S}^{(t)}$ has a nonnegative dominant right eigenvector $\mathbf v^{(t)}$ satisfying
$\mathbf{T}^{(t)}\mathbf{S}^{(t)}\mathbf v^{(t)}
=\rho(\mathbf{T}^{(t)}\mathbf{S}^{(t)})\mathbf v^{(t)}$ \cite{horn2012matrix}.
We use $v_i^{(t)}$ as a heuristic participation score and select the robot with the largest component.
For a power-cap violation, the culprit is the robot with the largest normalized requirement $\rho_i^{\star(t)}/\rho_{\max}$.
We denote these ordered tests and localization rules by \textsc{RrmCheck}$(\mathbf A^{(t)},\mathbf B^{(t)})$.
It returns $(\boldsymbol\rho^\star,d)$, with $d=\varnothing$ for a feasible allocation and $d=(\text{failure mode},i^\star)$ otherwise.
The complete trajectory is solver-verified only when Alg.~\ref{alg:rrm} returns a feasible allocation for every $t\in\mathcal T$.
Otherwise, the validator returns a failed slot, failure mode, and culprit robot to the trajectory planner.
Algorithm~\ref{alg:rrm} summarizes the per-slot allocation and validation procedure.

\begin{algorithm}[thb]
\caption{RRM validation for problem~\eqref{eq:RRM_subproblem}}\label{alg:rrm}
\begin{algorithmic}[1]
\REQUIRE $\{h_{i,b}^{(t)}\}$, $\{R_i\}$, $B$, $\rho_{\max}$, $N_0$, $K_{\mathrm{init}}$, and $L_{\max}$
\ENSURE Feasible $(\mathbf A^{(t)},\mathbf B^{(t)},\boldsymbol\rho^{(t)})$ or diagnostic $d$
\STATE $\mathcal S\leftarrow$ strongest-link seed and $K_{\mathrm{init}}$ restarts
\STATE Run \textsc{RrmCheck} for each seed after equal per-BS bandwidth initialization
\STATE \textbf{return} infeasible and a recorded $d$ \textbf{if every seed fails}
\STATE Initialize $(\mathbf A^{(t)},\mathbf B^{(t)},\boldsymbol\rho^{(t)})$ with the lowest-objective feasible seed
\FOR{$\ell=1,\dots,L_{\max}$}
    \STATE Generate $\mathbf A_{\mathrm{new}}^{(t)}$ by one-user reassignment
    \STATE Update $\mathbf B_{\mathrm{new}}^{(t)}$ at the two affected BSs using \eqref{eq:optimal_beta}
    \STATE $(\boldsymbol\rho_{\mathrm{new}}^\star,d_{\mathrm{new}})\leftarrow$ \textsc{RrmCheck}$(\mathbf A_{\mathrm{new}}^{(t)},\mathbf B_{\mathrm{new}}^{(t)})$
    \IF{$d_{\mathrm{new}}=\varnothing$ and $F(\mathbf A_{\mathrm{new}}^{(t)})<F(\mathbf A^{(t)})$}
        \STATE $(\mathbf A^{(t)},\mathbf B^{(t)},\boldsymbol\rho^{(t)})\leftarrow(\mathbf A_{\mathrm{new}}^{(t)},\mathbf B_{\mathrm{new}}^{(t)},\boldsymbol\rho_{\mathrm{new}}^\star)$
    \ENDIF
\ENDFOR
\STATE \textbf{return} feasible and $(\mathbf A^{(t)},\mathbf B^{(t)},\boldsymbol\rho^{(t)})$
\end{algorithmic}
\end{algorithm}

\myparagraph{Wireless-Conflict-Aware Multi-Robot Planning}
With the validator interface defined, we now describe how the planner generates and revises candidate trajectories.
For an instantiated problem $\problem[\mathsf{env},\mathsf{cfg}(\mathbf u)]$, the geometric subproblem is
\begin{align}
    \min_{\mathbf X}~&\sum_{i,t}\|\mathbf x_i^{(t+1)}-\mathbf x_i^{(t)}\|
    ~\text{s.t.}~\textbf{(TP Constr.)}.
    \label{eq:tp_subproblem}
\end{align}

\begin{figure}[t]
    \centering
    \includegraphics[width=\linewidth]{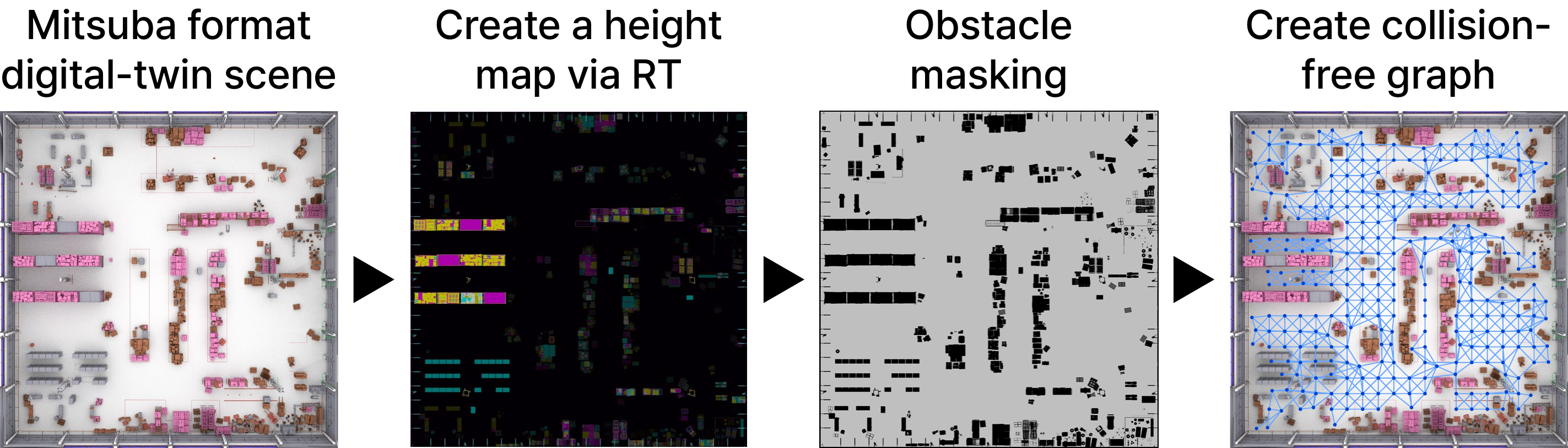}
    \caption{Collision-free navigation-graph construction from the DT scene.}
    \Description{A four-step pipeline that renders a top-down depth map from a Mitsuba-format scene, masks obstacles, and constructs a graph over the remaining navigable region.}
    \label{fig:02_collision_avoidance}
\end{figure}

Figure~\ref{fig:02_collision_avoidance} summarizes the navigation-graph construction.
The DT scene is rendered from above to obtain a depth map \cite{MITSUBA2}; thresholding the elevations produces a binary navigable mask.
Nodes are sampled with the clearance in \eqref{eq:constraint_safe_distance}, and candidate edges connect nodes reachable within the displacement bound in \eqref{eq:constraint_velocity}.
Bresenham's algorithm rejects edges that cross an obstacle \cite{Bresenham65-IBM}, after which isolated nodes and disconnected components are removed.

Standard CBS generates joint trajectories on this graph by detecting Type-I physical collisions and branching on space--time constraints for the involved robots \cite{Sharon12-AAAI,Tajbakhsh24-ICRA,Moldagalieva24-icra}.
Physical collision handling alone cannot detect paths whose simultaneous uplinks violate the radio constraints \cite{Kottinger22-IROS,Jiang26-IoTJ}.
Wireless CBS therefore augments the CBS conflict test with the RRM validator from Section~III-A.

For each candidate joint trajectory $\mathbf X$, the validator evaluates the induced channels over all slots.
If every slot is feasible, wireless CBS returns $\mathbf X$ and the corresponding $(\mathbf A,\mathbf B,\mathbf P)$.
Otherwise, it selects the earliest failed slot $t^\star$ and receives a culprit robot $i^\star$ from the interference-burst or power-cap diagnostic.
The point $(\mathbf x_{i^\star}^{(t^\star)},t^\star)$ is added to the CBS tree as a robot-specific space--time constraint, forcing the next candidate to avoid the diagnosed wireless conflict.
Thus, Type-I and Type-II conflicts are represented by the same constraint interface, closing the RRM--feasibility--replanning loop without changing the core CBS search logic.

\section{Data Lake and FastConfigNet}

Because \lucid continuously orchestrates TP--RRM optimization schemas based on dynamic operator intents, repeated deterministic trajectory-dependent RRM validation becomes a major computational bottleneck.
To support a responsive DITL loop, \lucid records solver evaluations in a data lake and uses FastConfigNet as a learning-based preconditioner.

\myparagraph{Data Lake: Archiving Evaluated Configurations}
Each solver call records the operator intent, the instantiated system configuration $\mathsf{cfg}(\mathbf u)$, the DT environment $\mathsf{env}$, the candidate trajectories, the optimized RRM variables, and the resulting solver outcome.
For candidates rejected by the validator, the record also includes the failure mode and localized culprit, such as the robot responsible for an interference burst or power-cap violation.
These records preserve the relation between intent, candidate operating point, scene-dependent channels, and solver outcomes, while providing training data for subsequent candidate screening.

\begin{figure}[htb]
    \centering
    \includegraphics[width=\linewidth]{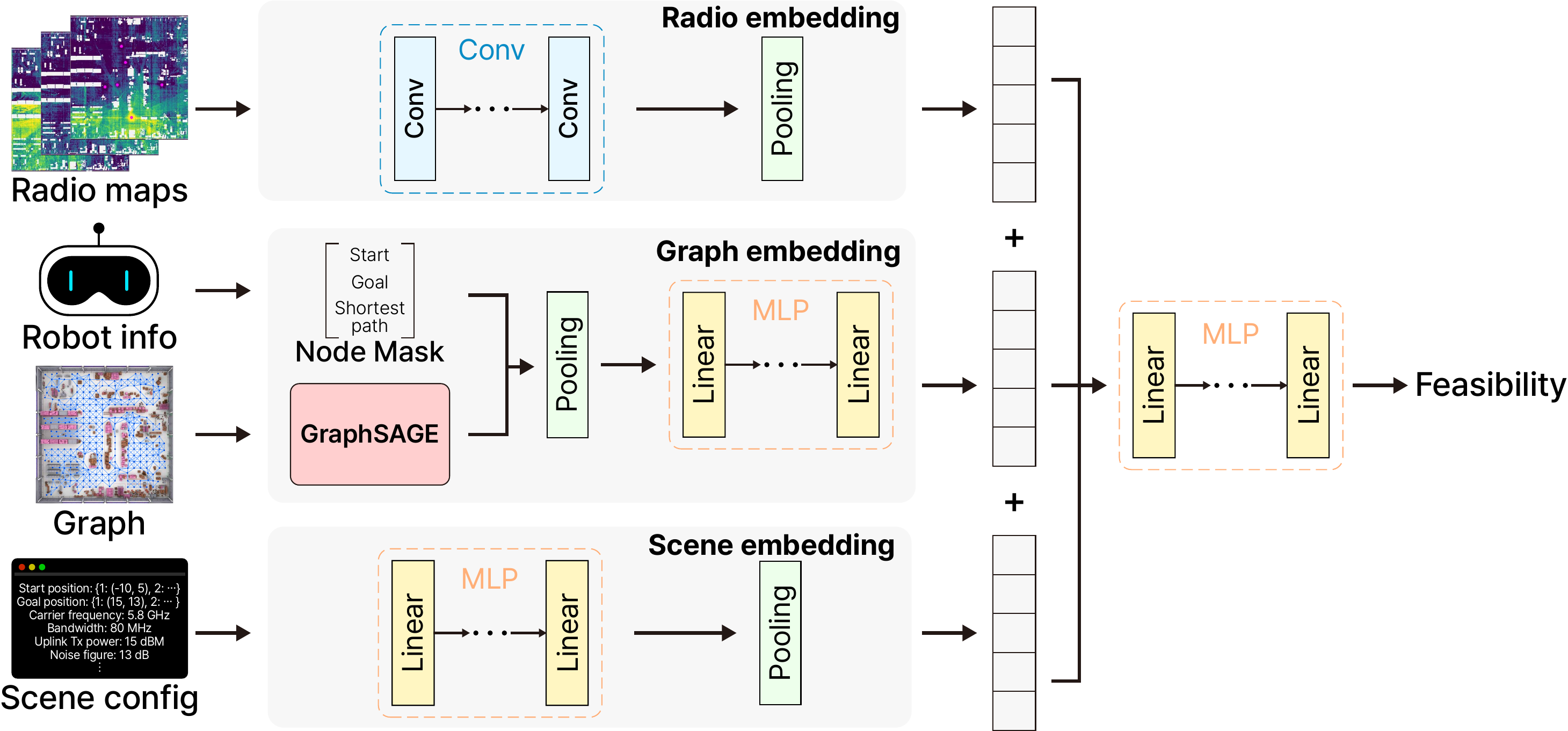}
    \caption{Multimodal encoding architecture of FastConfigNet.}
    \Description{A multimodal architecture that encodes radio maps, graph and robot information, and scene configuration before predicting candidate feasibility.}
    \label{fig:03_neural_net_architecture}
\end{figure}

\myparagraph{FastConfigNet: A Feasibility Surrogate Model}
FastConfigNet estimates the validator outcome of a candidate configuration before deterministic TP--RRM evaluation.
As illustrated in Fig.~\ref{fig:03_neural_net_architecture}, its multimodal encoder supports a varying number of robots and heterogeneous spatial inputs.
A DeepSets-based CNN maps the ray-tracing channel maps to a radio embedding that is invariant to the number of BSs \cite{Zaheer17-NIPS}.
In parallel, a GraphSAGE encoder represents the navigation graph and robot start, goal, and shortest-path information \cite{Hamilton17-GraphSage}.
These representations are combined with the global scene and system-configuration features in a multilayer prediction head.

\myparagraph{Model Training and Integration}
The prediction head estimates the candidate failure probability.
Training uses class-weighted binary cross-entropy to account for the predominance of infeasible samples in congested settings, a stratified train--validation--test split, and early stopping based on validation loss.

During intent-driven operating-point search, FastConfigNet estimates each candidate's feasibility probability to select a configuration that is likely to satisfy the operator's request.
For an operator who fixes 10 AMRs and seeks the highest supportable QoS, it forwards the highest-QoS configuration whose predicted feasibility probability exceeds a threshold to the TP--RRM solver.
Once the RRM validator confirms that configuration as feasible, the search terminates, avoiding lower-priority solver evaluations and reducing computation.

\section{SimBridge: From Robotics Scenes to Wireless-Ready DTs}

\begin{figure}[htb]
    \centering
    \includegraphics[width=\linewidth]{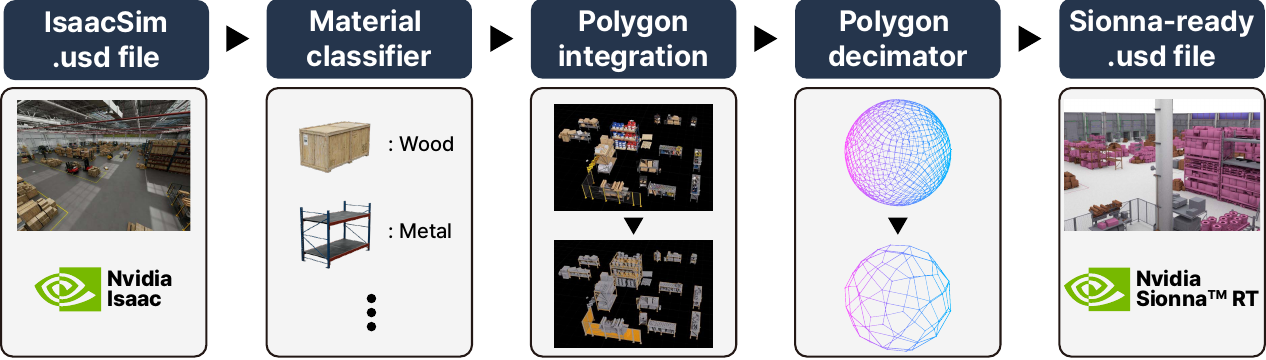}
    \caption{Scene adaptation pipeline from a source DT to a wireless-ready scene.}
    \Description{A five-stage pipeline of the evaluated implementation. An Isaac Sim USD scene is annotated with radio-material identifiers, its material-consistent meshes are integrated and simplified, and the resulting scene is exported as Mitsuba XML and binary PLY geometry for Sionna RT.}
    \label{fig:01_SimBridge_Preprocessor}
\end{figure}

SimBridge addresses two scene-construction obstacles identified in Sec.~\ref{sec:intro} by preparing a source robotics DT for trajectory-dependent propagation evaluation in \lucid.
Isaac Sim assets contain visual and physical descriptions but lack the radio-material identifiers required by Sionna RT, while their rendering-oriented geometry makes repeated propagation evaluation costly.
SimBridge takes a scene containing geometry, asset instances, and visual or physical metadata, and produces a material-aware, geometry-reduced wireless-ready representation for the propagation stage of the DITL environment.
In our implementation, SimBridge processes Isaac Sim USD assets, preserves radio-material boundaries during polygonal reduction, and exports the resulting scene as Mitsuba XML with binary PLY geometry for Sionna RT.
Figure~\ref{fig:01_SimBridge_Preprocessor} summarizes this processing flow.

\myparagraph{VLM-Assisted EM Material Assignment}
Robotics DT scenes are assembled by placing reusable assets, such as pallets, boxes, and racks, throughout the environment.
The full-scale warehouse used in our evaluation is assembled from 756 asset files.
To assign radio materials, SimBridge pairs textual metadata from each asset file, including its source path, with a $256\!\times\!256$ thumbnail.
These complementary textual and visual cues describe both the asset's semantic role and its appearance.
SimBridge supplies this text--image pair for each asset to Qwen2-VL-7B-Instruct \cite{Wang24-arXiv}.
The prompt restricts the model to a predefined set of ITU electromagnetic material categories \cite{ITUMaterials25-ITU}, and a canonicalization step converts the model response into the corresponding material identifier.
The resulting radio material attribute is propagated to every scene prim and instance that references the corresponding asset, covering 17,130 prims in the evaluated warehouse.

\begin{figure}[thb]
    \centering
    \includegraphics[width=0.9\linewidth]{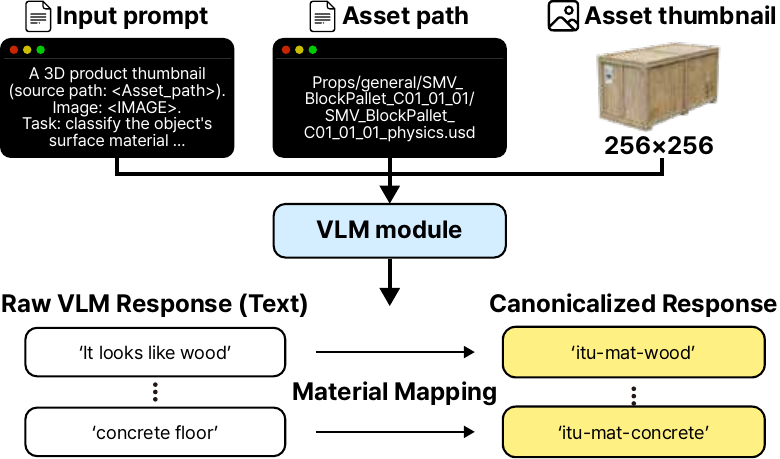}
    \caption{VLM-assisted mapping from asset metadata and a thumbnail to an ITU electromagnetic material identifier.}
    \Description{An asset's source-path text and representative thumbnail are supplied to a visual-language model. The model selects from a fixed set of ITU EM material categories, and the canonical identifier is assigned to all matching scene instances.}
    \label{fig:01_material_classifier}
\end{figure}
\begin{figure}[thb]
    \centering
    \includegraphics[width=0.9\linewidth]{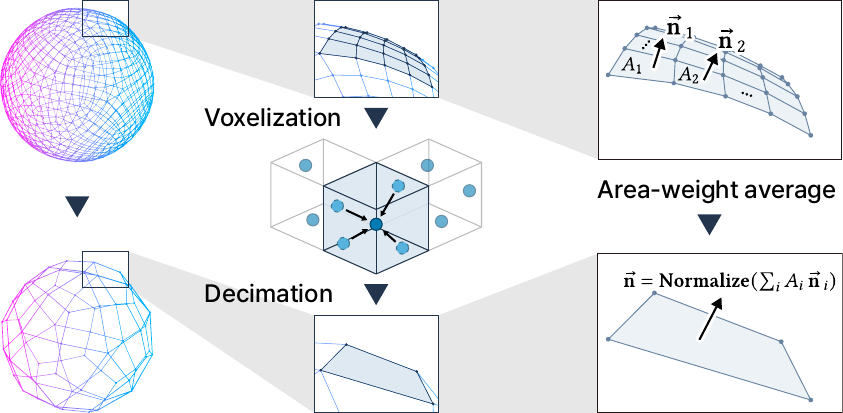}
    \caption{Material-preserving mesh simplification in SimBridge.}
    \Description{A dense input mesh is triangulated and its vertices are clustered according to spatial voxel and surface-normal direction. Invalid and duplicate faces are removed to produce a reduced mesh while retaining material boundaries.}
    \label{fig:01_polygon_decimator}
\end{figure}

\myparagraph{Material-Preserving Geometry Simplification}
Mesh simplification can reduce the complexity of detailed rendering assets \cite{OH25-CGF}.
The geometry-reduction stage reduces the cost of repeated propagation evaluation while preserving material assignments and coarse surface geometry.
SimBridge applies each instance's world transform to its mesh and aggregates the transformed geometry by ITU material identifier.
It then triangulates each mesh and merges vertices that fall within the same spatial voxel and quantized surface-normal bin, as illustrated in Fig.~\ref{fig:01_polygon_decimator}.
Vertices are clustered using their original surface normals. After clustering, nonmanifold faces are removed and the final vertex normals are recomputed from area-weighted face normals.
Area-weighted face normals matter because they let larger faces dominate the recomputed normal, preserving the true surface orientation after simplification — which in turn keeps ray incidence and reflection angles accurate.

For the full-scale warehouse, the scene contains 68,314,408 vertices and 89,481,477 faces and occupies 1.3~GB.
After the scene decimation, the resulting scene contains 12,196,481 vertices and 12,066,933 faces and occupies 350.6~MB.

\begin{figure}[t]
    \centering
    \includegraphics[width=\linewidth]{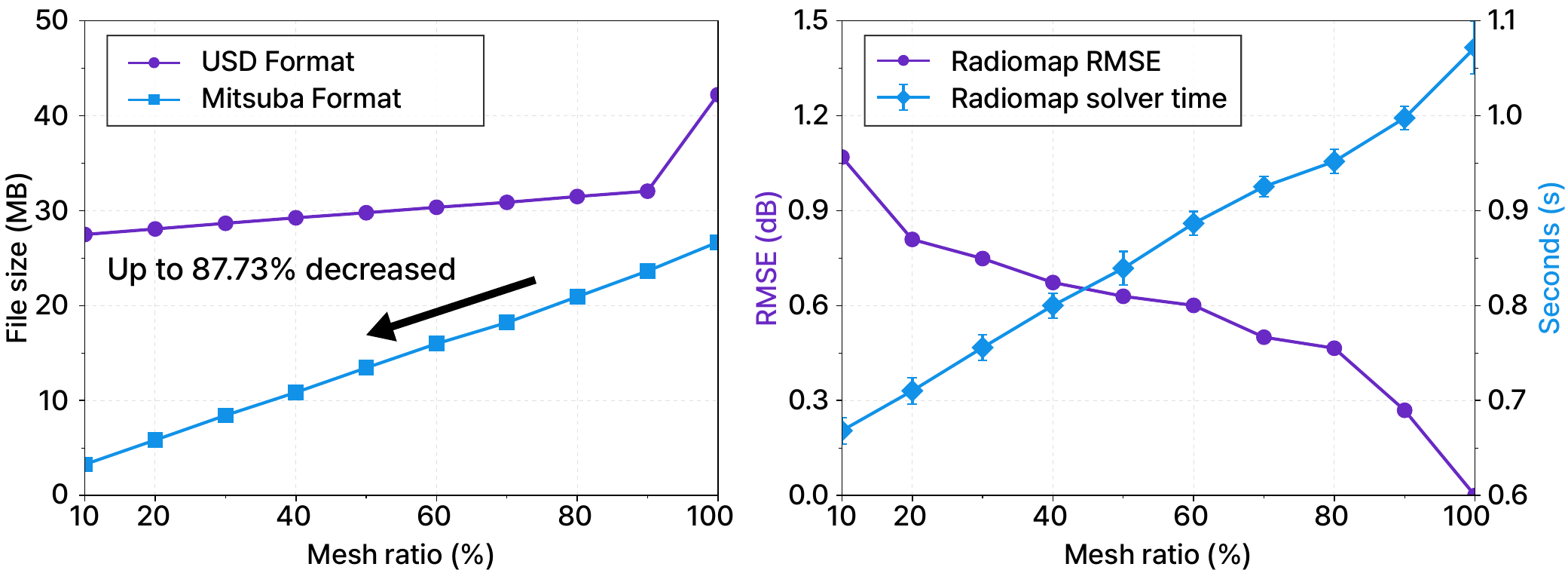}
    \caption{Mesh-reduction trade-off on the controlled small-warehouse scene. \textbf{Left}: USD and Mitsuba XML--PLY size. \textbf{Right}: path-gain RMSE relative to the 100\% mesh and scene-load-plus-solver time.}
    \label{fig:01_DT_decimate_ratio}
    \Description{Two plots show measurements as the retained mesh ratio increases from 10 to 100 percent. USD and Mitsuba XML plus PLY sizes increase with the ratio. Path-gain RMSE relative to the 100-percent mesh decreases from about 1.1 dB to zero, while the time from scene loading through path-gain tensor materialization increases from about 0.64 seconds to 1.07 seconds. Error bars summarize 50 trials.}
\end{figure}

We separately quantify the fidelity--cost trade-off on the warehouse scene.
We define the retained mesh ratio as the number of faces in the simplified mesh divided by that in the original mesh.
We vary this ratio from 10\% to 100\% and use the path-gain map generated from the unsimplified 100\% mesh as the reference.
For each ratio, we conduct 50 trials at 5.8~GHz, varying the ray-sampling random seed across trials, with a 0.5~m radio-map cell size and $4\times10^9$ ray samples per transmitter.
The reported runtime is measured from the start of scene loading until the EM propagation engine has produced and materialized the path-gain tensor.
As shown in Fig.~\ref{fig:01_DT_decimate_ratio}, lower retention mesh ratios reduce both representation size and end-to-end evaluation time, while increasing the path-gain error relative to the unsimplified mesh.
At a 10\% retention ratio, the RMSE is about 1.1~dB and the runtime is 0.64~s, compared with 1.07~s for the 100\% reference.

\section{Numerical Experiments}

In this section, we evaluate the performance and scalability of \lucid as a dynamic schema orchestrator.
Our experiments demonstrate how \lucid seamlessly unifies the active AMR count, QoS, and TP into a cohesive solution driven by evolving operator intent.
First, we evaluate the reliability of the interface that maps operator requests to TP--RRM configuration schemas.
We then test the end-to-end adaptability of the LLM-orchestrated framework in a multi-phase scenario, for which we deploy Gemini 3.6 Flash as the underlying LLM agent \cite{Google25-arXiv}.
All experiments run on a server equipped with an AMD Ryzen Threadripper 7970X CPU and two NVIDIA RTX 6000 Ada Generation GPUs.
The multi-phase experiment measures how the resulting pipeline adjusts operating points under these changes.
We also compare the RRM association search with a small-instance exhaustive oracle.
Finally, we evaluate FastConfigNet in terms of predictive accuracy, trajectory generation speed, and few-shot adaptation to new propagation environments.

\myparagraph{Intent-to-Configuration Reliability}
We evaluate how reliably different LLM backends map an operator's natural-language request to the intended TP--RRM configuration schema. This requires more than extracting numerical parameters: the model must determine which fields of $\mathrm{cfg}(u)$ are fixed, searched, or relaxed when a request is infeasible. As shown in Table~\ref{tab:operator_intent_stress}, Claude Sonnet 5 and Gemini 3.6 Flash achieve the highest overall schema-selection accuracy at 94\% and 91\%, followed by GPT-4.1 mini at 83\%. GPT-4.1 nano and Llama 3.1 8B achieve 62\% and 39\%, showing a clear gap in mapping complex operator priorities and relaxation rules to the intended optimization schema.

\begin{table}[t]
\centering
\caption{Reliability of LLM backends for operator requests.}
\label{tab:operator_intent_stress}
\setlength{\tabcolsep}{2.5pt}
\begin{adjustbox}{width=\columnwidth}
\begin{tabular}{l ccc cc}
\toprule
\multirow{2}{*}{Backend / Model}
& \multicolumn{3}{c}{Reliability and safety}
& \multicolumn{2}{c}{Stress categories} \\
\cmidrule(lr){2-4}
\cmidrule(lr){5-6}
& \shortstack{Overall\\acc. ($\uparrow$)}
& \shortstack{Raw validator\\rejection ($\downarrow$)}
& \shortstack{Semantic\\escape ($\downarrow$)}
& \shortstack{Injection\\acc. ($\uparrow$)}
& \shortstack{Ambig./conflict\\acc. ($\uparrow$)} \\
\midrule
\arrayrulecolor{lightgray}
Claude Sonnet 5 & \textbf{94.0} & \textbf{0.0} & \textbf{6.0} & \textbf{95.0} & \textbf{80.0} \\
\cmidrule(l{2pt}r{2pt}){1-1}\cmidrule(l{2pt}r{2pt}){2-6}
Gemini 3.6 Flash & 91.0 & 1.0 & 9.0 & 80.0 & \textbf{80.0} \\
\cmidrule(l{2pt}r{2pt}){1-1}\cmidrule(l{2pt}r{2pt}){2-6}
GPT-4.1 mini & 83.0 & 1.0 & 16.0 & 50.0 & 72.0 \\
\cmidrule(l{2pt}r{2pt}){1-1}\cmidrule(l{2pt}r{2pt}){2-6}
GPT-4.1 nano & 62.0 & 60.0 & 7.0 & 50.0 & 44.0 \\
\cmidrule(l{2pt}r{2pt}){1-1}\cmidrule(l{2pt}r{2pt}){2-6}
Llama 3.1 8B & 39.0 & 72.0 & 15.0 & 35.0 & \textbf{80.0} \\
\arrayrulecolor{black}
\bottomrule
\end{tabular}
\end{adjustbox}
\vspace{-5mm}
\end{table}

The detailed metrics further distinguish the reliability of the backends. The raw validator rejection rate remains at only 0--1\% for Claude Sonnet 5, Gemini 3.6 Flash, and GPT-4.1 mini, but rises sharply to 60--72\% for the smaller backends, indicating frequent structurally or physically invalid configurations. Semantic escape remains nonzero across all models, ranging from 6\% to 16\%, showing that structurally valid outputs can still misrepresent the intended schema. The stronger backends also perform more reliably under prompt injection and ambiguous or conflicting requests, although their accuracy is not perfect in these stress cases. Overall, stronger backends more consistently select the intended schema while producing fewer validator-rejected outputs. We next evaluate how this guarded intent-to-configuration interface supports end-to-end adaptation under changing operator requests.

\begin{figure}[htb]
    \centering
    \includegraphics[width=\linewidth]{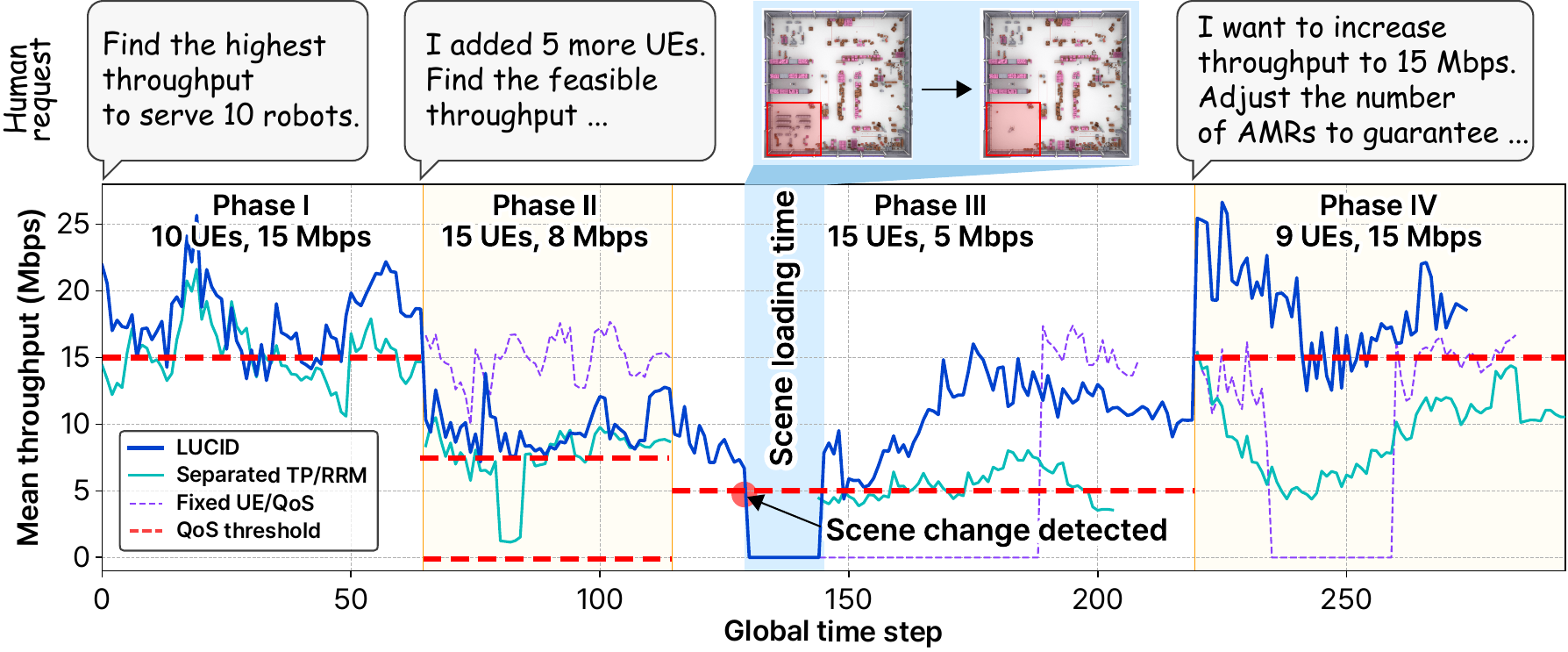}
    \caption{Performance of \lucid under changes in operator intent, active AMR count, and physical environment.}
    \Description{A multi-phase timeline comparing LUCID with separated TP/RRM and fixed-configuration baselines as the requested AMR count, QoS, and scene change over time.}
    \label{fig:03_multi_phase_scenario}
\end{figure}
\begin{figure*}
    \centering
    \includegraphics[width=\linewidth]{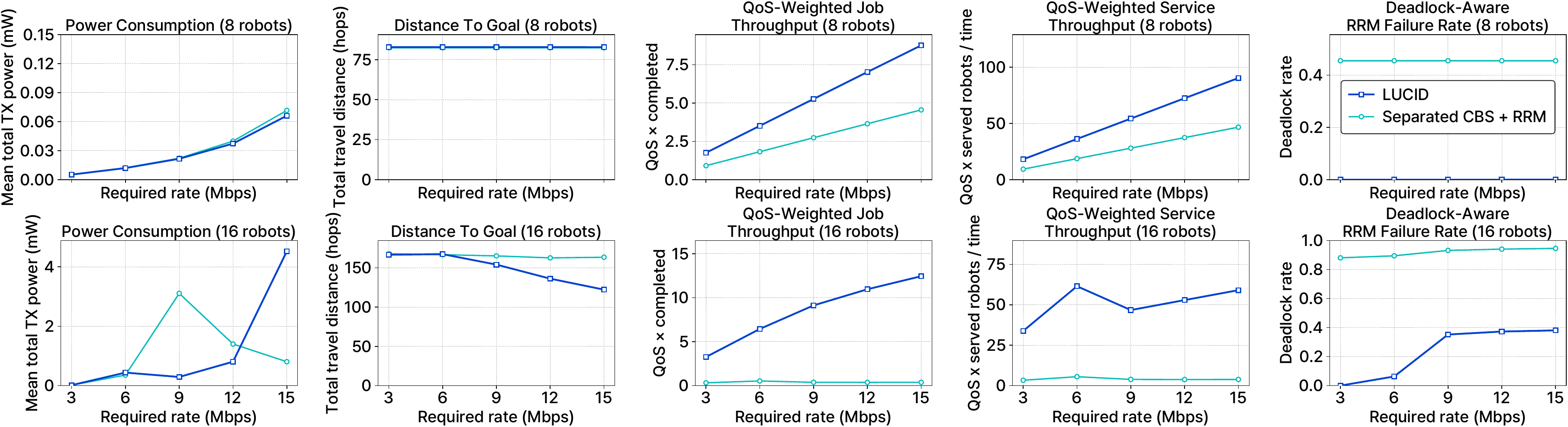}
    \caption{End-to-end comparison of \lucid and \textbf{Separated TP/RRM}.}
    \label{fig:comparison with separated TP/RRM}
    \Description{
    Ten line plots compare power consumption, distance to goal, QoS-weighted job throughput, QoS-weighted service throughput, and deadlock-aware RRM failure rate for 8- and 16-robot settings. Each plot varies the required communication rate from 3 to 15 Mbps and compares LUCID with Separated CBS + RRM.
    }
\end{figure*}

\myparagraph{Multi-phase scenarios}
Figure \ref{fig:03_multi_phase_scenario} evaluates the end-to-end adaptability of \lucid under changes in operator intent, active AMR count, and physical environmental conditions.

We evaluate \lucid against the two baseline approaches:
\begin{itemize}[leftmargin=*]
    \item \textbf{Separated TP/RRM} performs RRM after executing a vanilla CBS-based TP, adjusting the allocation to meet the required QoS throughput.
    The number of operating robots and the QoS throughput are assigned identically to those used by \lucid in each phase.
    \item \textbf{Fixed UE/QoS} assumes a conventional static configuration and operates 10 robots with a fixed throughput requirement of 15 Mbps regardless of the phase.
\end{itemize}
Below is the performance comparison across phases:
\begin{itemize}[leftmargin=*]
    \item \textbf{Phase I (Initial state)}: The system begins with an operator's request to deploy 10 UEs with the highest possible throughput.
    All schemes successfully orchestrate trajectories and radio resources to maintain AMR uplink QoS above the threshold.

    \item \textbf{Phase II (Workload scaling)}: The operator requests to add 5 more robots to the scene.
    As the environment is too congested to support 15 UEs at the previous QoS requirement of 15 Mbps, \lucid dynamically searches for a new feasible operating point.
    Prioritizing the active AMR count over individual data rates as requested, the agent seamlessly reconstructs the optimization schema to degrade the required throughput to a feasible 8 Mbps, allowing all 15 robots to operate safely.

    \item \textbf{Phase III (Environmental Change)}: The physical environment changes through the removal of cargo boxes or changes in rack positions.
    SimBridge converts the changed scene into a Sionna-compatible representation, after which the DT runtime recomputes the ray-tracing channel maps before TP--RRM reevaluation.
    Baselines without a scene update capability fail to find solutions that satisfy the QoS requirements as the overall wireless propagation environment degrades.
    In contrast, \lucid continues to serve the 15-AMR deployment by further lowering the QoS threshold.

    \item \textbf{Phase IV (QoS change)}: The new request from the operator now prioritizes throughput over the number of UEs, requesting 15 Mbps of throughput while adjusting the active AMR count accordingly.
    Here, \lucid shifts the problem template to fix the QoS requirement and optimize the active AMR count.
    Both baselines fail to guarantee QoS in deep fading regions, whereas \lucid ensures QoS by adaptively adjusting the trajectories.
\end{itemize}

Throughout the timeline, \lucid consistently satisfies the dynamically shifting QoS threshold.
In contrast, the baseline methods experience severe performance drops and deadline misses because they treat TP and RRM as isolated, static problems.
They cannot jointly adapt physical routing, the active AMR count, and wireless constraints to match the evolving operator intent.

Figure~\ref{fig:comparison with separated TP/RRM} provides a detailed comparison of the end-to-end behavior of the two designs. Under moderate load (8 robots), \lucid consistently outperforms Separated TP/RRM across all communication rates. The QoS-weighted job throughput is approximately 1.9$\times$ higher throughout the evaluated range, and the QoS-weighted service throughput shows a similar improvement. While \lucid maintains a goal-reaching ratio of 1.0, Separated TP/RRM reaches only about half of the robots on average. By contrast, the average travel distance and transmit power remain nearly identical between the two methods, indicating that under moderate contention the improvement comes from converting comparable resources into substantially more completed tasks.

The difference becomes qualitatively stronger under heavy load (16 robots). \lucid preserves a goal-reaching ratio of 1.0 across all tested rates, whereas Separated TP/RRM drops to a near-zero completion regime. As a result, the QoS-weighted job throughput differs by more than one order of magnitude, and the QoS-weighted service throughput exhibits a similarly large separation. The deadlock-aware RRM failure rate combines observed RRM-infeasible slots with the penalty assigned to deadlocked trials; it ranges from 0.00 to 0.38 for \lucid, compared with 0.88 to 0.95 for Separated TP/RRM across the evaluated requirements. These results indicate that wireless-aware joint planning prevents the low-completion operating regime that arises in the separated pipeline under high spatial contention.

The power and distance results support the same interpretation. For 8 robots, both methods consume comparable transmit power, showing that the gain of \lucid is not obtained by spending more power. For 16 robots, Separated TP/RRM can consume higher power without translating that expenditure into useful service. As the communication rate increases, the average travel distance of \lucid decreases, consistent with selectively committing to QoS-feasible robots and completing their tasks efficiently.

\begin{figure}[htb]
    \centering
    \begin{minipage}[t]{0.48\linewidth}
        \centering
        \includegraphics[width=\linewidth]{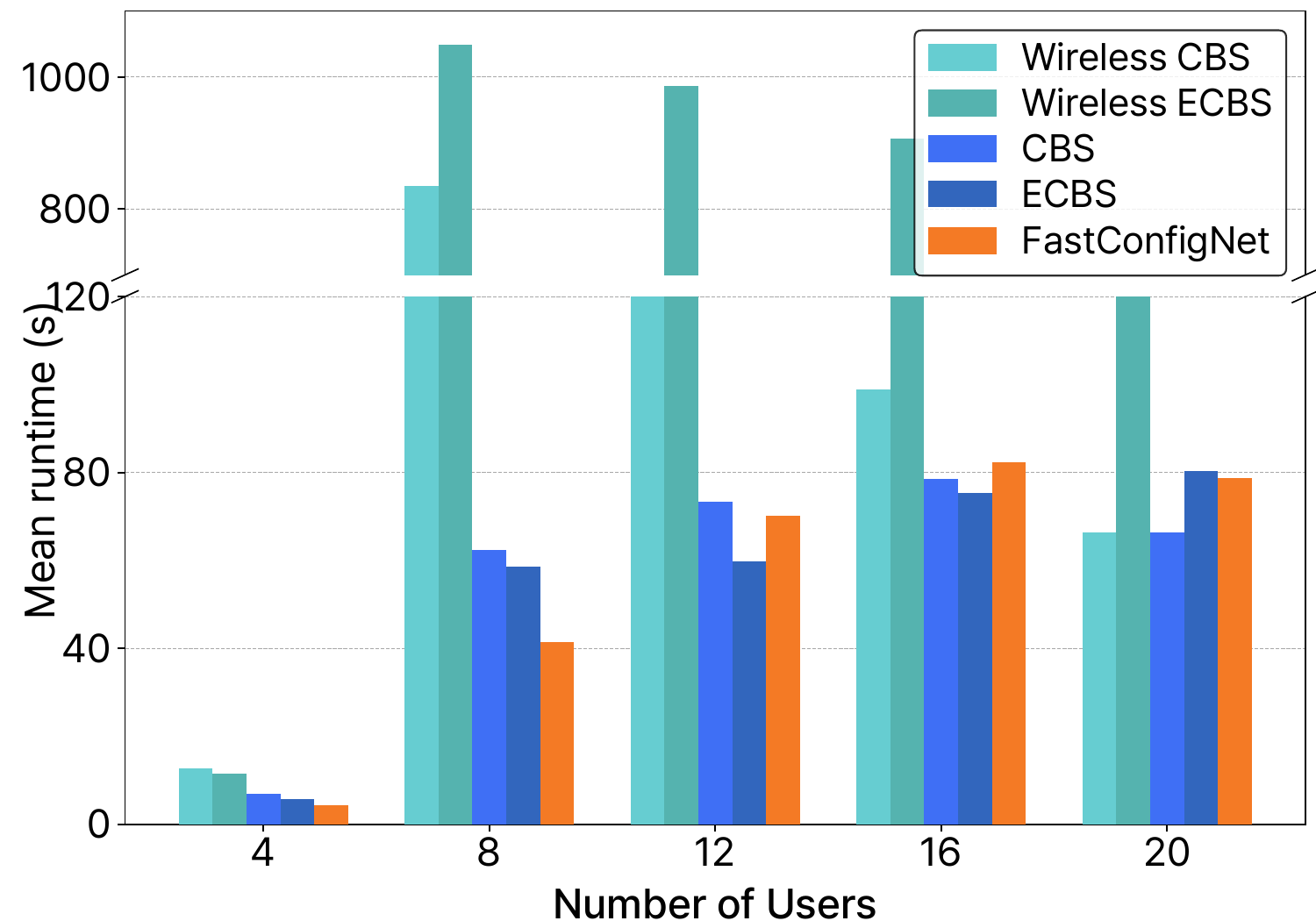}
        \refstepcounter{figure}\label{fig:03_runtime_sec_mean_bar}
        \footnotesize{Fig.~\thefigure.} Mean compute runtime to determine the feasibility of TP.
        \Description{A grouped bar chart comparing the "Mean runtime (ms)" on the y-axis for trajectory generation across five different algorithms as the "Number of Users" on the x-axis increases (4, 8, 12, 16, and 20). The y-axis features a scale break, separating the lower range (0 to 120 ms) from the upper extreme range (800 to over 1000 ms). The five algorithms are Wireless CBS (light blue), Wireless ECBS (teal), CBS (blue), ECBS (dark blue), and NN (orange). At 4 users, all methods run very quickly, well under 20 ms. However, at 8 users, the runtime for Wireless ECBS spikes dramatically to over 1000 ms, and Wireless CBS jumps to over 800 ms. For 12, 16, and 20 users, Wireless ECBS remains extremely high (around 900 to 1000 ms), while Wireless CBS drops back down to the 60-120 ms range. Throughout the increases in users, the other three methods (CBS, ECBS, and NN) remain consistently low and stable, gradually increasing but staying comfortably under 100 ms.}
    \end{minipage}\hfill
    \begin{minipage}[t]{0.48\linewidth}
        \centering
        \includegraphics[width=\linewidth]{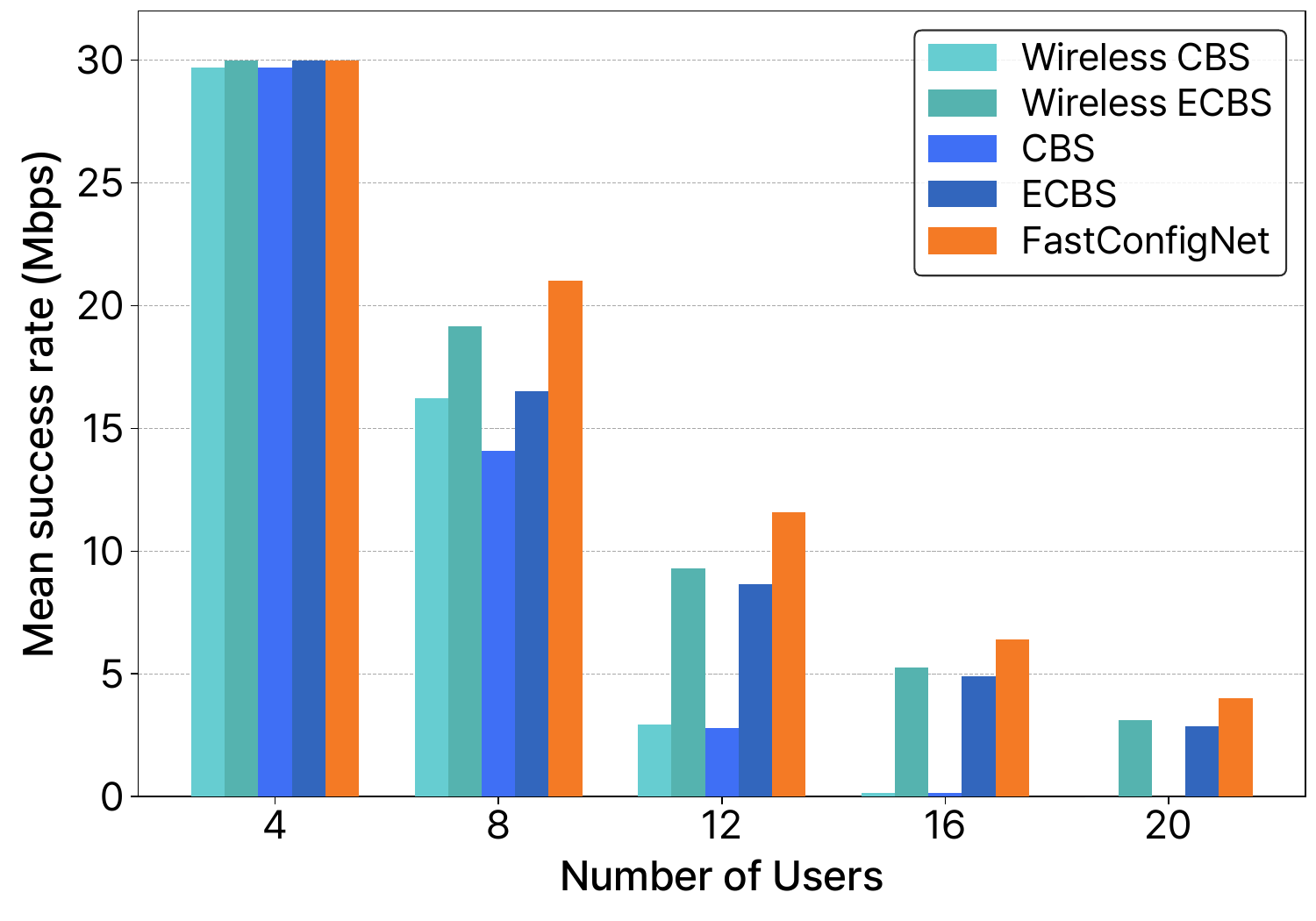}
        \refstepcounter{figure}\label{fig:03_success_rate_mbps_mean_bar}
        \footnotesize{Fig.~\thefigure.} Maximum achievable uplink QoS.
        \Description{A grouped bar chart evaluating the mean runtime in milliseconds against the number of users (4, 8, 12, 16, and 20) for five different methods: Wireless CBS, Wireless ECBS, CBS, ECBS, and NN. Overall, the runtime for all methods starts near 30 ms at 4 users and exhibits a clear downward trend as the user count increases. Among the evaluated methods, the NN approach consistently shows the highest runtime across all user groups, followed closely by Wireless ECBS and ECBS. In contrast, CBS and Wireless CBS consistently achieve the lowest runtimes. While NN and the ECBS variants maintain runtimes between 3 and 6 ms at 16 and 20 users, the runtimes for CBS and Wireless CBS drop to near-zero values at these stages.}
    \end{minipage}
\end{figure}

\myparagraph{Performance of FastConfigNet}
We evaluate FastConfigNet against four distinct pathfinding baselines: Wireless CBS, Wireless ECBS, vanilla CBS, and vanilla ECBS.\footnote{ECBS is a bounded-suboptimal variant of the standard CBS algorithm used for multi-agent pathfinding \cite{Barer21-SoCS}, which allows slightly longer paths in order to significantly reduce the number of nodes that must be explored.}
The comparative results are detailed in Figs. \ref{fig:03_runtime_sec_mean_bar} and \ref{fig:03_success_rate_mbps_mean_bar}.

Figure \ref{fig:03_runtime_sec_mean_bar} shows the average runtime needed to generate multi-robot trajectories as the number of AMRs increases from 4 to 20.
Baseline geometric planners, CBS and ECBS, maintain low execution times in all scenarios because the search space only handles physical space-time conflicts.
In contrast, the link-aware methods, Wireless \{CBS, ECBS\}, introduce significant computational overhead that grows rapidly as the environment becomes more crowded.
As the AMR count grows, link-aware planners discard more paths that fail to satisfy throughput constraints. This reduces the effective search space and leads to shorter runtimes.
FastConfigNet closes the performance gap by using fast feed-forward inference to screen candidate trajectories before deterministic RRM validation.
This screening reduces the number of expensive validator calls while retaining deterministic verification for the selected solution.

Figure \ref{fig:03_success_rate_mbps_mean_bar} shows the maximum achievable uplink throughput per robot as the AMR count increases.
With 4 users, all algorithms reach nearly the maximum target data rate.
As the number of robots grows, the environment becomes more congested, and the achievable throughput decreases for all methods.
At the evaluated AMR counts, Wireless \{CBS, ECBS\} and FastConfigNet achieve higher per-robot throughput than the geometric planners, CBS and ECBS.
This gap arises because CBS and ECBS do not account for wireless link capacity limits, which leads to Type-II wireless conflicts. In contrast, link-aware methods proactively route robots to avoid strong interference and power cap violations.
FastConfigNet's throughput trend indicates that its screening scores prioritize wireless-feasible trajectories before the retained solution undergoes deterministic RRM validation.

\begin{figure}[htb]
    \centering
    \includegraphics[width=\linewidth]{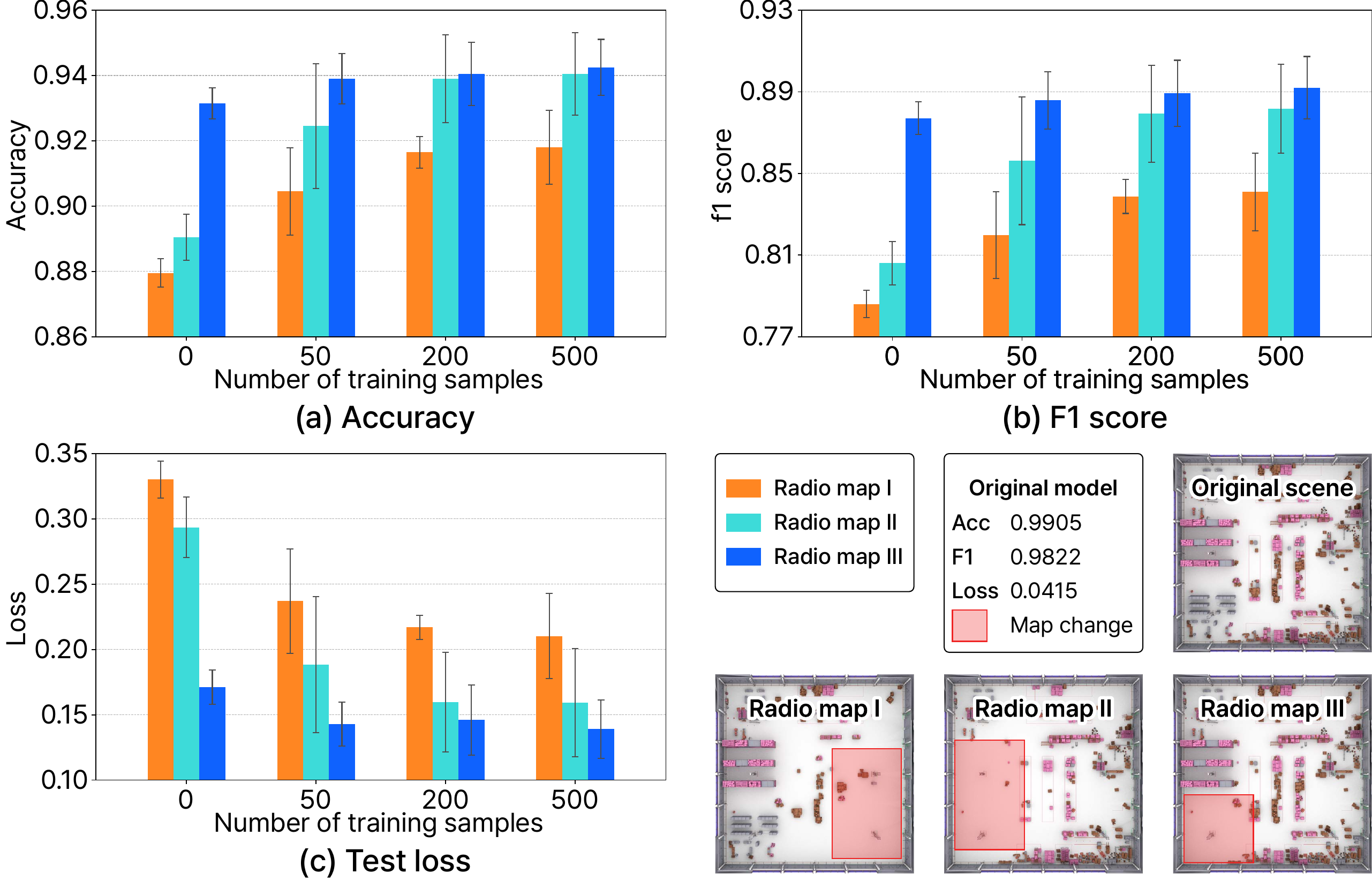}
    \caption{Few-shot learning for scene adaptation.}
    \label{fig:03_nn_adaptation}
    \Description{A composite figure with four panels demonstrating few-shot learning for scene adaptation. Panels (a), (b), and (c) are bar charts showing Accuracy, F1 score, and Test loss, respectively, on the y-axes against the Number of training samples (0, 50, 200, and 500) on the x-axes. Each x-axis category has three bars representing three different modified environments: Radio map I (orange), Radio map II (light blue), and Radio map III (dark blue). Across all three charts, as the number of training samples increases from 0 to 500, the Accuracy and F1 scores steadily increase, while the Test loss steadily decreases, indicating successful adaptation. The bottom-right section provides a legend and visual context. A text box shows the "Original model" baseline metrics (Acc: 0.9905, F1: 0.9822, Loss: 0.0415). Next to it are four top-down warehouse map thumbnails: an "Original scene" and three modified scenes labeled "Radio map I, II, and III". In each modified scene, a red shaded rectangle of varying size indicates the specific area of "Map change" (physical layout alterations) that the model is adapting to.}
\end{figure}

Figure~\ref{fig:03_nn_adaptation} illustrates how the proposed model adapts to scene changes using only a small number of additional training samples.
When the propagation environment changes, the original model trained on the previous scene initially experiences a noticeable drop in accuracy and F1 score. However, by fine-tuning with a limited number of new samples, the model rapidly recovers its performance.
As the number of adaptation samples increases, both accuracy and F1 score steadily improve while the test loss decreases, indicating effective learning of the new environment. These results demonstrate that the model can quickly adapt to unseen propagation conditions through few-shot retraining, enabling practical deployment in dynamically changing scenes.

\section{Conclusion}
This paper presented \lucid, an agentic AI framework that shifts cloud-robotics TP--RRM from solving fixed optimization instances to dynamically orchestrating optimization problem schemas.
Rather than treating trajectory planning and radio resource management as isolated or rigid problems, \lucid interprets their formulation as a predefined template and dynamically configures its variables, objectives, and constraints according to evolving operator intent and scene context.
The LLM orchestrates this schema-level configuration, while a deterministic wireless-aware planner and RRM solver perform the optimization and feasibility verification.
To enable such orchestration at scale, SimBridge converts large-scale Isaac Sim scenes into material-aware, geometry-reduced wireless DTs suitable for repeated DITL evaluation, while the Data Lake records previous evaluations and FastConfigNet screens new candidate configurations before deterministic validation.
The evaluated warehouse scenarios show how these components adapt feasible operating points as the requested AMR count, QoS priorities, and scene conditions change.
Ultimately, the results demonstrate that dynamic, intent-driven TP--RRM schema orchestration is feasible in a large-scale DITL testbed, overcoming the fixed-formulation and testbed limitations that have constrained prior cloud-robotics research.

\bibliographystyle{IEEEtran}
\bibliography{references.bib}

\end{document}